\documentclass[twocolumn,twocolappendix]{aastex}		

\usepackage{natbib}

\shorttitle{The PMS of Westerlund 2}
\shortauthors{Zeidler et al.}

\begin{document}

\title{A High-Resolution Multiband Survey of Westerlund 2 With the \textit{Hubble Space Telescope}. II. Mass accretion in the Pre-Main Sequence Population}

\author{Peter~Zeidler\altaffilmark{1,2}, Eva~K.~Grebel\altaffilmark{1}, Antonella~Nota\altaffilmark{2,3},  Elena~Sabbi\altaffilmark{2},  Anna~Pasquali\altaffilmark{1}, Monica~Tosi\altaffilmark{5}, Alceste~Z.~Bonanos\altaffilmark{4}, \and Carol~Christian\altaffilmark{2}}

\altaffiltext{1}{Astronomisches Rechen-Institut, Zentrum f\"ur Astronomie der Universit\"at Heidelberg, M\"onchhofstr. 12-14, 69120 Heidelberg, Germany (\texttt{pzeidler@ari.uni-heidelberg.de})}
\altaffiltext{2}{Space Telescope Science Institute, 3700 San Martin Drive, Baltimore, MD 21218, USA}
\altaffiltext{3}{ESA, SRE Operations Devision}
\altaffiltext{4}{IAASARS, National Observatory of Athens, GR-15326 Penteli, Greece}
\altaffiltext{5}{INAF - Osservatorio Astronomico di Bologna}

\begin{abstract}
We present a detailed analysis of the pre-main-sequence (PMS) population of the young star cluster Westerlund~2 (Wd2), the central ionizing cluster of the \ion{H}{2} region RCW~49, using data from a high resolution multi-band survey with the \textit{Hubble} Space Telescope. The data were acquired with the Advanced Camera for Surveys in the $F555W$, $F814W$, and $F658N$ filters and with the Wide Field Camera 3 in the $F125W$, $F160W$, and $F128N$ filters. We find a mean age of the region of $1.04\pm0.72$~Myr. The combination of dereddened $F555W$ and $F814W$ photometry in combination with $F658N$ photometry allows us to study and identify stars with H$\alpha$ excess emission. With a careful selection of 240 bona-fide PMS H$\alpha$ excess emitters we were able to determine their H$\alpha$ luminosity, which has a mean value $L(\rm{H}\alpha)=1.67 \cdot 10^{-31}~\rm{erg}~\rm{s}^{-1}$. Using the PARSEC 1.2S isochrones to obtain the stellar parameters of the PMS stars we determined a mean mass accretion rate $\dot M_{\rm{acc}}=4.43 \cdot 10^{-8}~M_\odot~\rm{yr}^{-1}$ per star. A careful analysis of the spatial dependence of the mass-accretion rate suggests that this rate is $\sim 25\%$ lower in center of the two density peaks of Wd2 in close proximity to the luminous OB stars, compared to the Wd2 average. This rate is higher with increasing distance from the OB stars, indicating that the PMS accretion disks are being rapidly destroyed by the far-ultra-violet radiation emitted by the OB population.
\end{abstract}


\section{Introduction}
\label{sec:introduction}
With a stellar mass of M~$\ge 10^4$~M$_\odot$ \citep{Ascenso_07} the young Galactic star cluster \object{Westerlund~2} \citep[hereafter Wd2;][]{Westerlund_61} is one of the most massive young clusters in the Milky Way (MW). It is embedded in the \ion{H}{2} region \object{RCW~49} \citep{Rodgers_60}, located in the Carina-Sagittarius spiral arm $(\alpha,\delta)=(10^h23^m58^s.1,-57^\circ45'49'')$(J2000), $(l,b)=(284.3^\circ,-0.34^\circ)$.

There is general agreement in the literature that Wd2 is younger than 3~Myr and that its core might be younger than 2~Myr \citep{Ascenso_07,Carraro_13}. In our first paper \citep[][hereafter Paper I]{Zeidler_15} we confirmed the cluster distance of \citet{Vargas_Alvarez_13} of 4.16~kpc, using \textit{Hubble} Space Telescope (HST) photometry and our high-resolution 2D extinction map. We estimated the age of the cluster core to be between 0.5 and 2.0~Myr. Using two-color diagrams (TCDs), we found a total-to-selective extinction $R_V=3.95 \pm 0.135$ (Paper I). This value was confirmed by an independent, numerical study of \citet{Mohr-Smith_15}. Their best-fitting parameter is $R_V=3.96^{+0.12}_{-0.14}$, which is in very good agreement with our result. Furthermore, we found that Wd2 contains a rich population of pre-main-sequence (PMS) stars.

Over the past decades studies showed that during the PMS phase, low-mass stars grow in mass through accretion of matter from their circumstellar disk \citep[e.g.,][and references therein]{Lynden-Bell_74,Calvet_00}. These disks form due to the conservation of angular momentum following infall of mass onto the star, tracing magnetic field lines connecting the stars and their disks. It is believed that this infall leads to the strong excess emission in the infrared in contrast to the flux distribution of a normal black-body. This excess emission is observed for many PMS stars and probably originates through gravitational energy being radiated away and exciting the surrounding gas. As a result, this excess can be used to measure accretion rates for these classical T-Tauri stars \citep[especially via H$\alpha$ and Pa$\beta$ emission lines, e.g.,][]{Muzerolle_98c,Muzerolle_98b}. The accretion luminosity ($L_{acc}$) can then be used to calculate the mass accretion rate ($\dot M$). Studies of different star formation regions \citep[e.g., Taurus, Ophiuchus,][]{Sicilia-Aguilar_06} showed that these accretion rates decrease steadily from $\sim 10^{-8} \rm M_\odot \rm{yr}^{-1}$ to less than $10^{-9} \rm M_\odot \rm{yr}^{-1}$ within the first 10~Myr of the PMS star lifetime \citep[e.g.,][]{Muzerolle_00,Sicilia-Aguilar_06}. This is in good agreement with the expected evolution of viscous disks as described by \citet{Hartmann_98}. These studies all agree that the mass accretion rate decreases with the stellar mass.

Understanding these accretion processes plays an important role in understanding disk evolution as well as the PMS cluster population as a whole \citep{Calvet_00}. The ''standard'' way to quantify the mass accretion is through spectroscopy. Usually, one studies the intensity and profile of emission lines such as H$\alpha$, Pa$\beta$, or Br$\gamma$, which requires medium- to high-resolution spectra. This approach has the disadvantage of long integration times and, therefore, only a small number of stars can usually be observed.

H$\alpha$ filters have long been used to identify H$\alpha$ emission-line objects in combination with additional broadband or intermediate-band colors \citep[e.g.,][]{Underhill_82}. For panoramic CCD detectors, the technique was first applied by \citet{Grebel_92} and then developed further for different filter combinations and to quantify the H$\alpha$ emission \citep[e.g.,][]{Grebel_93b,Grebel_97}. \citet{deMarchi_10} used this photometric method to estimate the accretion luminosity of PMS stars. Normally the R-band is used as the continuum for the H$\alpha$ filter. \citet{deMarchi_10} showed for the field around SN~1987A \citep{Romaniello_98,Panagia_00,Romaniello_02} that the Advanced Camera for Surveys \citep[ACS,][]{ACS} filters $F555W$ and $F814W$ can be similarly used to obtain the continuum for the H$\alpha$ filter. Up to now, this method \citep{deMarchi_15} has been proven to be successful in studies for different clusters, such as NGC~346 in the Small Magellanic Cloud \citep[SMC, ][]{deMarchi_11a} and NGC~3603 in the MW \citep{Beccari_10}.

Due to its young age, Wd2 is a perfect target to study accretion processes of the PMS stars in the presence of a large number \citep[$\sim 80$, see][]{Moffat_91} of O and B stars. In close proximity to OB stars, the disks may be expected to be destroyed faster by the external UV radiation originating from these massive stars. This would lead to a lower excess of H$\alpha$ emission in the direct neighborhood of the OB stars \citep{Anderson_13,Clarke_07}. Our high-resolution multi-band observations of Wd2 in the optical and near-infrared (Paper I) give us the opportunity to study the PMS population and the signatures of accretion in detail in a spatially resolved, cluster-wide sample down to a stellar mass of 0.1~M$_\odot$. In Paper I, we showed that the stellar population of RCW~49 mainly consists of PMS stars and massive OB main-sequence (MS) stars. These objects are not found in one single, centrally concentrated cluster but are mostly located in two sub-clusters of Wd2, namely its main concentration of stars, which we term the ''main cluster'' (MC), and a secondary, less pronounced concentration, which we call the ''northern clump'' (NC).

This paper is a continuation of the study presented in Paper I with an emphasis on the characterization of the PMS population. In Sect. \ref{sec:catalog} we give a short overview of the photometric catalog presented in Paper I. In Sect. \ref{sec:stellar_pop} we look in more detail into the stellar population of RCW~49. We analyze the color-magnitude diagrams (CMDs) for the region as a whole as well as for individual sub-regions. In Sect. \ref{sec:Halpha} we provide a detailed analysis of the determination of the H$\alpha$ excess emission stars. In Sect. \ref{sec:acc_L_and_M} we use H$\alpha$ excess emission to derive the accretion luminosity as well as the mass accretion rate. Furthermore, we provide a detailed analysis of the change of the mass accretion rate with the stellar age and the location relative to the OB stars. In Sect.~\ref{sec:uncertainties} we give an overview and summary of the contribution of the different sources of uncertainty. In Sect. \ref{sec:summary} we summarize the results derived in this paper and we discuss how they further our understanding of this region.

\section{The photometric catalog}
\label{sec:catalog}

\begin{figure*}[htb]
\resizebox{\hsize}{!}{\includegraphics{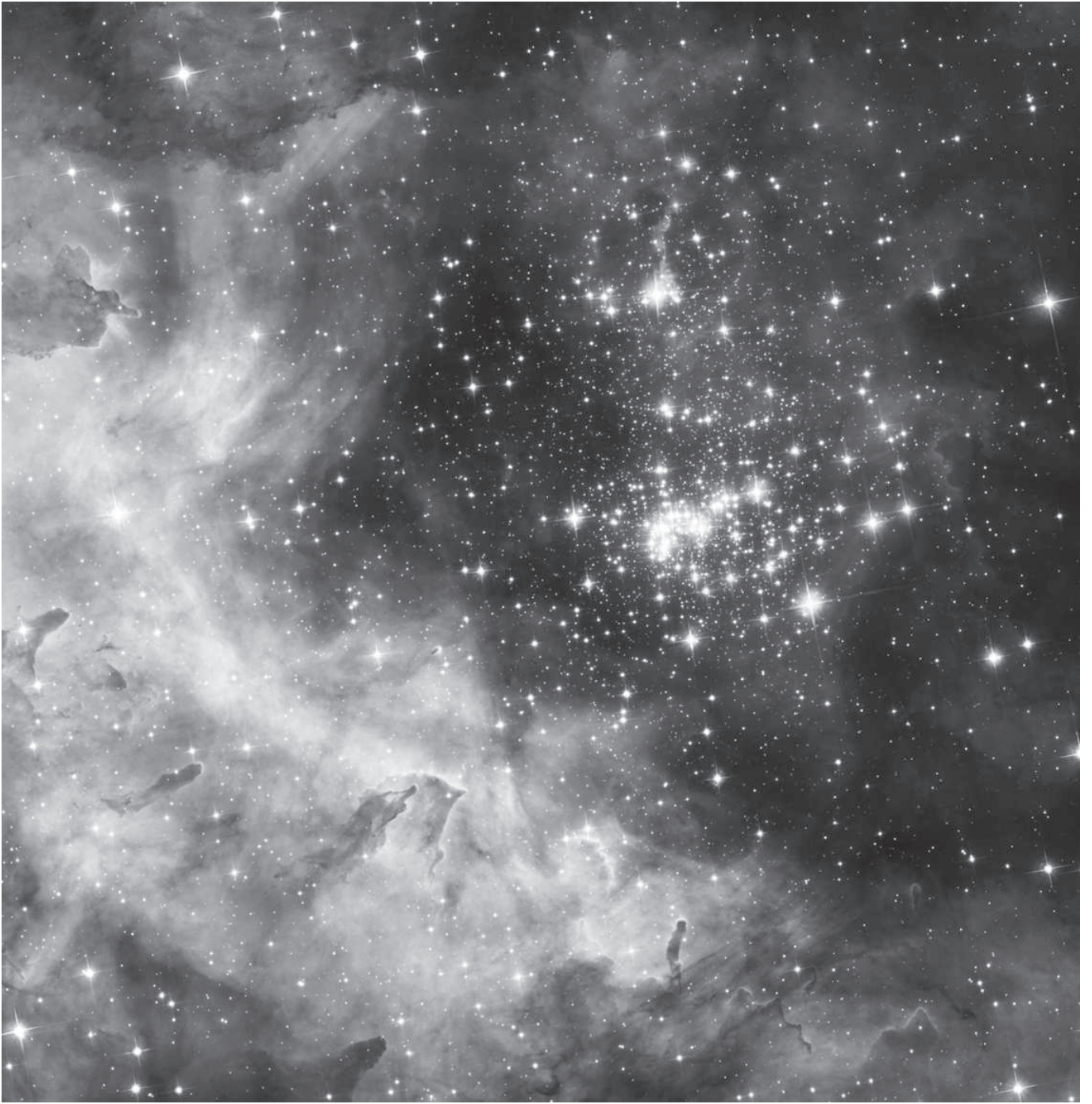}}
\caption{Black and white composite image of the HST ACS and WFC3/IR data of Wd2. A color image is presented in Paper~I (Fig.~2). North is up, East to the left. The FOV is $\sim$4~arcmin$\times$4~arcmin. The color image was chosen to be the official Hubble 25th anniversary image. Credit: NASA, ESA, the Hubble Heritage Team (STScI/AURA), A. Nota (ESA/STScI), and the Westerlund 2 Science Team.}
\label{fig:RGB_JIV}
\end{figure*}

The observations of Wd2 were performed with HST during Cycle 20 using the ACS and the IR Channel of the Wide Field Camera 3 \citep[WFC3/IR,][]{WFC3}. In total, six orbits were granted and the science images were taken on 2013 September 2 to 8 (proposal ID: 13038, PI: A. Nota). A detailed description of the observations, the data reduction, and the creation of the photometric catalog can be found in Paper I.

Wd2 was observed in four wide-band filters (ACS: $F555W$ and $F814W$, exposure times 1400~s; WFC3/IR: $F125W$ and $F160W$; exposure times 947~s). 3~s short exposures were included for the $F555W$ and $F814W$ filters to recover most of the saturated objects. Observations were also taken in two narrow-band filters (ACS: $F658N$, exposure times 1400~s and WFC3/IR: $F128N$, 748~s), centered on the H$\alpha$ and Pa$\beta$ line emission. The final catalog contains 17,121 objects that were detected in at least two filters. 2236 point sources were detected in all six filters. 90\% of all sources have a photometric error less than $\sigma_{F555W}=0.160$~mag, $\sigma_{F658N}=0.185$~mag, $\sigma_{F814W}=0.093$~mag, $\sigma_{F125W}=0.035$~mag, $\sigma_{F128NW}=0.095$~mag, and $\sigma_{F160W}=0.054$~mag. Our optical data are $\sim5$~mag deeper than the photometric data used by \citet{Vargas_Alvarez_13} obtained with the Wide-Field Planetary Camera 2 \citep{WFPC2}. Our near-infrared data are 3--5~mag deeper than the images obtained by \citet{Ascenso_07}. Our data were chosen to be the 25th anniversary image of the HST\footnote{\url{http://hubblesite.org/newscenter/archive/releases/2015/12/image/a/}}. A black and white version of the image is shown in Fig. \ref{fig:RGB_JIV}.

Using the $F658N$ and $F128N$ filters \citep{Pang_11,Zeidler_15} we were able to create a high-resolution pixel-to-pixel (0.098 arcsec pixel$^{-1}$) $E(B-V)_g$ color-excess map of the gas. Using the zero-age main sequence (ZAMS) derived from the \textsc{Padova and Trieste Stellar Evolution Code}\footnote{\url{http://stev.oapd.inaf.it/cmd}} \citep[hereafter: PARSEC 1.2S,][]{Bressan_12} with a Solar metallicity of $Z=0.0152$ \citep{Caffau_11} in combination with spectroscopic observations of the brightest stars of Wd2 \citep{Rauw_07,Rauw_11,Vargas_Alvarez_13}, we transformed the spatially resolved gas excess map into a stellar color excess with a median value of $E(B-V)_\star=1.55$~mag \citep[see Sect.~5.1 in][]{Zeidler_15}. This map was then used to deredden individual photometric measurements in our catalog.

Using two-color diagrams (TCDs), we found a value for the total-to-selective extinction of $R_V=3.95 \pm 0.135$ (see Paper I), using the extinction law of \citet{Cardelli_89}. This agrees with the range of $R_V$ values of 3.64-3.85 found in multiple studies of Wd2 \citep{Rauw_07,Rauw_11,Vargas_Alvarez_13,Hur_15}. From the spectral-energy distribution (SED) fitting of O and B-type stars observed with the VLT Survey Telescope (VST) \citet{Mohr-Smith_15} recently derived $R_V=3.96^{+0.12}_{-0.14}$, in excellent agreement with our finding. Plotting PARSEC 1.2S isochrones over CMDs and fitting the turn-on (TO) region where PMS stars join the MS we were able to confirm for Wd2 the distance $d=4.16$~kpc (Paper~I) as estimated by \citet{Vargas_Alvarez_13}.

Throughout this paper, unless stated differently, we will use $d=4.16$~kpc and $R_V=3.95 \pm 0.135$. All colors and magnitudes flagged with the subscript ''0'' were dereddened individually using the method described in Sect. 5 of Paper I. We revised the transformation law of the color excess $E(B-V)$ for the $F555W$ filter to better fit the TCDs. This is described in detail in the Appendix~\ref{sec:F555W_calib} and is used from now on.

\begin{figure*}
	\resizebox{\hsize}{!}{\includegraphics{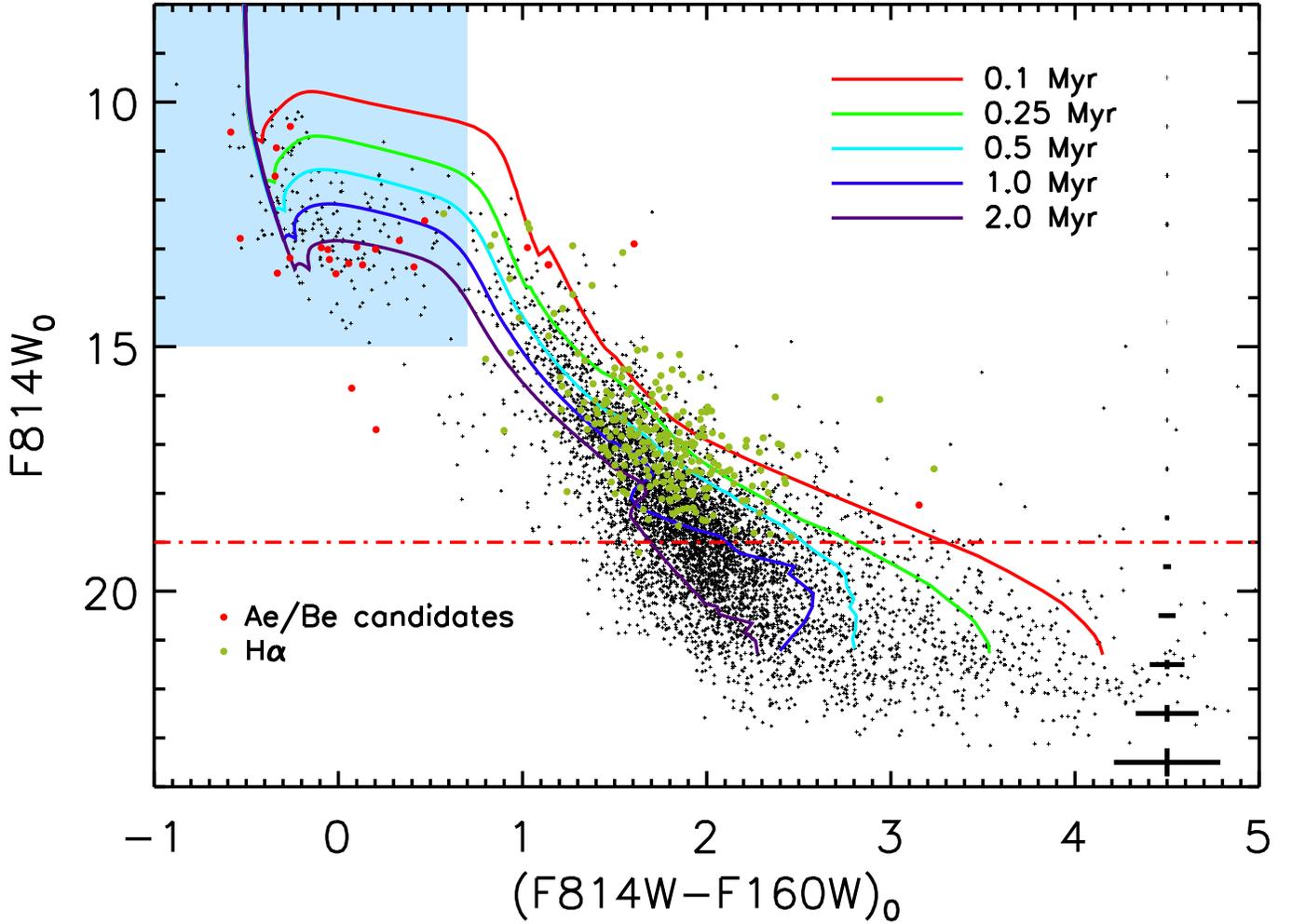}}
	\caption{The $F814W_0$ vs. $(F814W-F160W)_0$ CMD of the RCW~49 members (black dots). We defined all objects brighter than $F814W_0=15.0$~mag and bluer than $(F814W-F160W)_0=0.7$~mag as MS and TO members (light blue area). The red dots mark the 24 Ae/Be candidates, namely all stars showing H$\alpha$ excess and having an EW$>10\rm{\AA}$ but being bluer than $F555W-F814W=0.2$~mag. The green dots show all H$\alpha$ excess PMS objects. Overplotted are the PARSEC 1.2S isochrones \citep{Bressan_12} between 0.1 and 2.0~Myr for a Solar metallicity of $Z=0.0152$ and a distance of $d=4.16$~kpc. The red dash-dotted line marks the detection limit set by the $F555W$ and $F658N$ filters ($\sim 0.3~\rm{M}_\odot$ at an age of 1~Myr), both of which are needed to identify the H$\alpha$ excess stars. On the right side typical photometric uncertainties are shown as a function of magnitude.}
	\label{fig:F814W-F160W_F814W}
\end{figure*}

\section{The stellar population of RCW~49 - distribution and age}
\label{sec:stellar_pop}

To investigate the PMS population in more detail, we defined all objects brighter than $F814W_0=15.0$~mag and bluer than $(F814W-F160W)_0=0.7$~mag as a member of the main-sequence (MS) or TO region (see Fig.~\ref{fig:F814W-F160W_F814W}). This selection leaves us with 5404 PMS and 200 MS and TO objects.

We use different selection criteria for different samples of PMS stars:

\begin{itemize}
	\item For the 5404 PMS star candidates selected in Paper I (using their loci in the CMD), we require detection in both the $F814W$ and $F160W$ filters (from now on denoted as ''full-sample'' PMS stars).
	\item H$\alpha$ excess emission sources need to be detected in the $F555W$, $F814W$, $F160W$, and $F658N$ filters and to have an H$\alpha$ excess (green dots in Fig.~\ref{fig:F814W-F160W_F814W}).
	\item Because the $F555W$ and $F658N$ images are less deep than $F814W$ and $F160W$ we selected 1690 PMS stars from the full sample with the same detection criteria as our H$\alpha$ excess emission stars. This means they have to be probable cluster members and need to be detected in the $F555W$, $F814W$, $F160W$, and $F658N$ filters. From now on they are denoted as our ''reduced-sample'' PMS stars. These stars do not necessarily have H$\alpha$ excess emission.
\end{itemize}

The full-sample PMS stars is used for the properties of the Wd2 cluster and the RCW~49 region, while the reduced sample is always used to compare the H$\alpha$ excess-emitting stars with the non-emitting stars. The limiting magnitude is $F814W_0 \approx 19$~mag, which corresponds to a $\sim 0.3~\rm{M}_\odot$ star at an age of 1~Myr (red dash-dotted line in Fig.~\ref{fig:F814W-F160W_F814W}). We selected 240 H$\alpha$ excess sources (green dots in Fig.~\ref{fig:F814W-F160W_F814W}). The detailed determination, as well as the mass accretion rates are demonstrated in Sect.~\ref{sec:Halpha}.

\begin{figure}[t]
	\resizebox{\hsize}{!}{\includegraphics{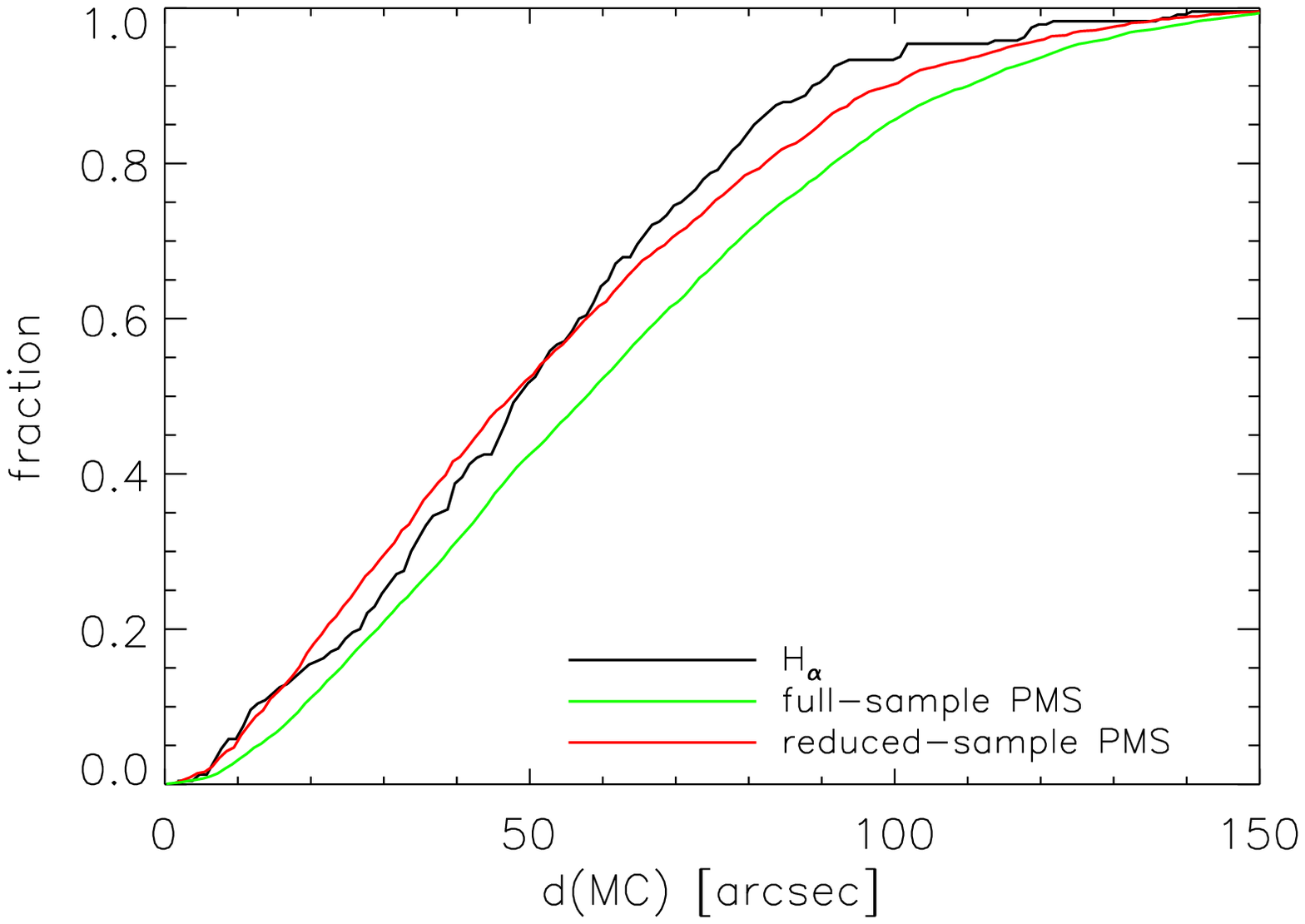}}
	\caption{The cumulative radial distribution in distance of the objects from the peak coordinates of the MC $(\alpha,\delta)=(10^h24^m02^s.4,-57^\circ45'33.44''$; J2000), determined with a 2D Gaussian fit \citep{Zeidler_16c}. At a distance of 4.16~kpc from the Sun $50''$ correspond to 1~pc.}
	\label{fig:cumulative_distribution}
\end{figure}

In Fig.~\ref{fig:cumulative_distribution} we plotted the cumulative distributions of the radial distance of the full-sample PMS, the reduced-sample PMS, and the H$\alpha$ excess emission sources. The coordinates of the central density peak of the MC \citep[$(\alpha,\delta)=(10^h24^m02^s.4,-57^\circ45'33.44'')$(J2000),][]{Zeidler_16c} were used as origin. A Kolmogorov-Smirnov (K-S) test yields a probability of only $\sim 11\%$ that the H$\alpha$ excess sources and the full-sample PMS share the same radial distribution, while it yields a $\sim 74\%$ probability that the H$\alpha$ excess sources and the reduced-sample PMS have the same radial distribution. This test and the distribution itself (see Fig.~\ref{fig:cumulative_distribution}) confirm that for comparing the stars with H$\alpha$ excess emission to the cluster members, the reduced sample of PMS stars needs to be used.

\subsection{A closer look at the PMS ages}
\label{sec:stellar_ages}
In Paper I we suggested an upper age limit of 2~Myr for the whole cluster. In this section we compare the age distribution of the reduced sample of PMS stars in Wd2. In Tab.~\ref{tab:ages} we list all 240 H$\alpha$ excess objects and the 1690 reduced-sample PMS stars for different ages. For comparison we also list the 5404 full-sample PMS stars.

\begin{figure}[t]
	\resizebox{\hsize}{!}{\includegraphics{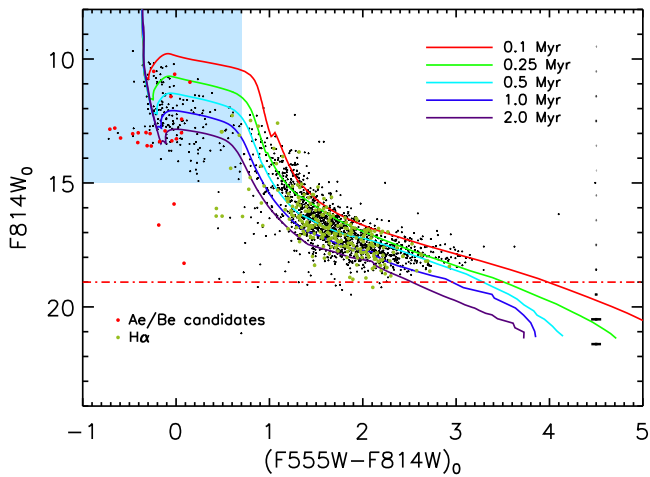}}
	\caption{The $F814W_0$ vs. $(F555W-F814W)_0$ CMD of the RCW~49 members. The remaining description is equivalent to Fig.~\ref{fig:F814W-F160W_F814W}.}
	\label{fig:F555W-F814W_F814W}
\end{figure}

The age distribution of what we call the reduced-sample PMS stars shows a mean age of $0.84 \pm 0.64$~Myr, while the mean age of the stars with H$\alpha$ excess is $0.62\pm0.57$~Myr. In comparison the full-sample PMS stars have a mean age of $1.04 \pm 0.71$~Myr. The difference in age between the full sample and the reduced sample most likely originates from the requirement that the latter additionally needs to be detected in the $F555W$ filter, which is less deep than the other filters (Paper~I, \cite{Zeidler_16c}). This argument is supported  by the $F814W_0$ vs. $(F555W-F814W)_0$ CMD (see Fig.~\ref{fig:F555W-F814W_F814W}). The slope of the locus of PMS stars in the CMD becomes shallower for lower masses. Therefore, younger stars can be detected down to lower masses than older stars since they are more luminous in these filters. This leads to the effect that the reduced sample (as well as the H$\alpha$ excess stars) have a younger mean age. We conclude that the age estimate from the full sample ($1.04 \pm 0.71$~Myr) better represents the age of the Wd2 region. It is in good agreement with the age of 1.5--2~Myr determined by \citet{Ascenso_07} and is in agreement with the MS lifetime of O3--O5 stars of $\sim2-5$~Myr \citep[see Tab 1.1,][]{Sparke_07}. The locus of the H$\alpha$ excess stars (green dots in Fig~\ref{fig:F814W-F160W_F814W}) appears to be slightly shifted to younger ages. This effect, additionally to the above described effect, is caused by a lower mass-accretion rate for older stars, resulting in a lower H$\alpha$ excess rate.
With these ages, Wd2 appears to be of the same age or even younger than the massive star cluster \object{HD97950} in the giant HII region \object{NGC~3603} \citep{Pang_13}, which has an age of about 1~Myr, \object{Trumpler~14} \citep[$\le 2$~Myr,][]{Carraro_04_Tr14} in the Carina Nebula \citep{Smith_08}, \object{Arches} \citep[$\sim2$~Myr,][]{Figer_02,Figer_05}, \object{R136} in the \object{LMC} \citep[1--4~Myr,][]{Hunter_95,Walborn_97,Sabbi_12,Sabbi_16}, and younger than \object{Westerlund~1} \citep[$5.0\pm1.0$~Myr,][]{Clark_05,Gennaro_11,Lim_13}.

\subsection{The individual regions in RCW~49}
\label{sec:regions_in_RCW49}
\begin{figure*}[htb]
	\resizebox{\hsize}{!}{\includegraphics{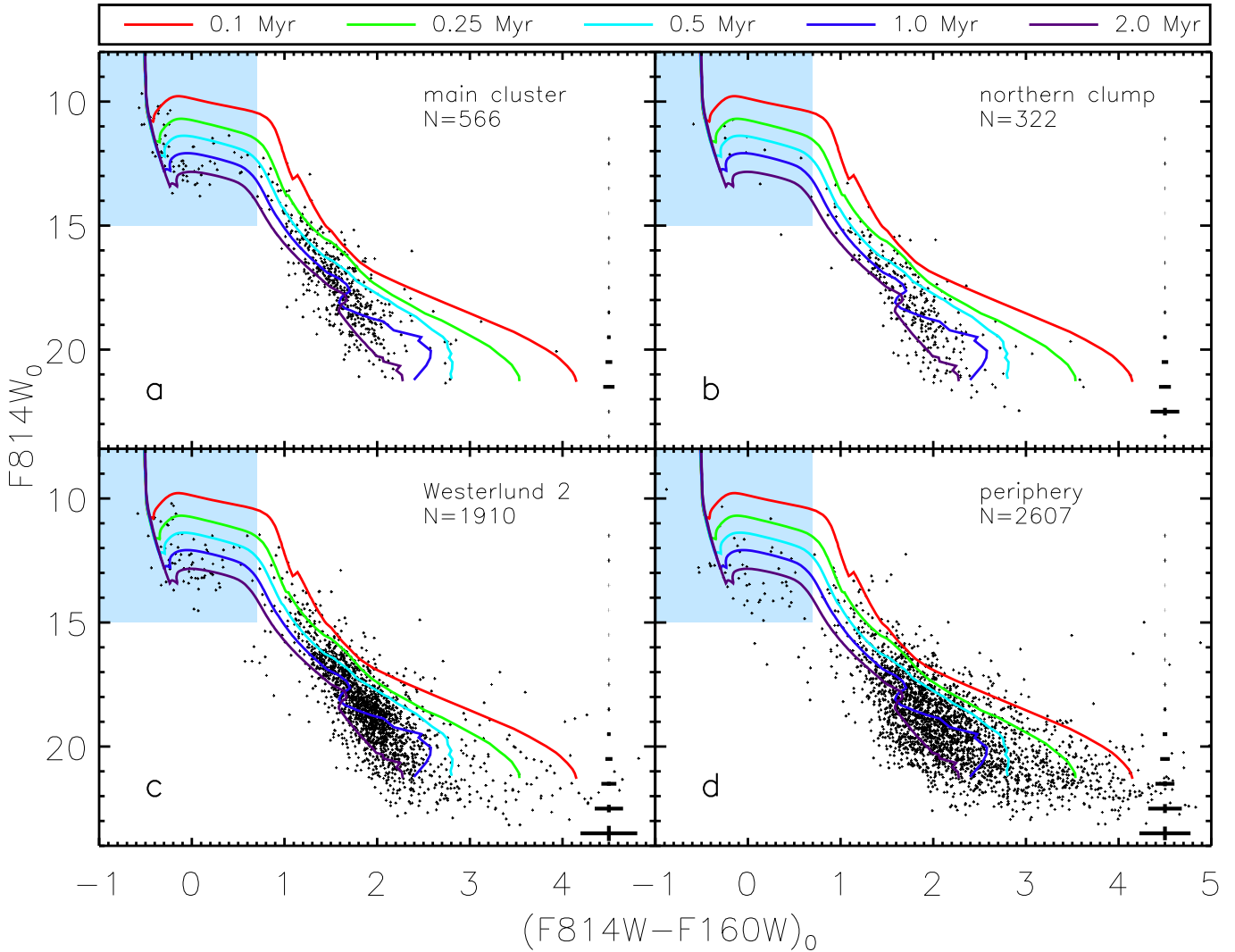}}
	\caption{The $F814W_0$ vs. $(F814W-F160W)_0$ CMDs of the four different regions of RCW~49.  All objects brighter than $F814W_0=15.0$~mag and bluer than $(F814W-F160W)_0=0.7$~mag are defined as MS or TO members (light blue area). Overplotted are the PARSEC 1.2S isochrones \citep{Bressan_12} between 0.1 and 2.0~Myr for a Solar metallicity of $Z=0.0152$ and a distance of $d=4.16$~kpc. On the right side typical photometric uncertainties are shown.}
	\label{fig:area_CMDs}
\end{figure*}

The stellar surface density map \citep{Zeidler_16c} of the RCW~49 region shows that this region can be divided in four parts: the MC, the NC, the remaining parts of the Wd2 cluster (1-$\sigma$ contour of the density profile excluding the MC and NC), and the Wd2 periphery. These regions are defined by a fit of two 2D Gaussian distributions with a common offset to the completeness-corrected stellar surface-density map of the RCW~49 member stars. We show a detailed analysis in \citet{Zeidler_16c}, which is more sophisticated than the one used in Paper~I. In Fig.~\ref{fig:area_CMDs} we show the $F814W_0$ vs. $(F814W-F160W)_0$ CMDs for each subregion. In the following Section we will analyze the distribution and properties of the different areas.

While we focus in Tab.~\ref{tab:ages} on the number of PMS stars per sample for each age bin, in Tab.~\ref{tab:spatial_distribution} we focus on the mean properties of the four different regions.

The MC hosts a well-populated MS, TO, and PMS. We selected 498 full-sample PMS members. The full-sample PMS stars define an age of $1.04 \pm 0.71$~Myr. The uncertainties are represented by the standard deviation of the ages. The 263 PMS stars of the reduced sample show a younger estimated age of $0.84 \pm 0.64 $~Myr, while the 36 H$\alpha$ stars located in the area of the MC have an estimated age of $0.69 \pm 0.57$~Myr (see Tab.~\ref{tab:spatial_distribution}).  The lack of very faint objects (compared to the other three regions) is caused by crowding and incompleteness effects \citep{Zeidler_16c}.

The NC hosts 310 full-sample PMS members. The full-sample PMS members lead to an age estimate of $1.16 \pm 0.67 $~Myr and thus are coeval with the MC. Also their age distribution is similar to that in the MC (see Tab.~\ref{tab:ages}). The NC hosts in total 26 H$\alpha$ excess stars with a mean age of $0.76 \pm 0.60 $~Myr.

The Wd2 cluster shows an extended halo \citep[$2\sigma$ boundary,][]{Zeidler_16c} around the MC and NC. At least 1814 objects in this region are defined PMS with the same mean age as the MC and NC. The MC and NC are excluded from this region. The 106 H$\alpha$ excess stars have an age of $0.60 \pm 0.56 $~Myr.

2752 full-sample PMS members are found in the periphery of RCW~49. Most of the objects in this region are fairly faint and red (compared to the distribution in the other three areas). With a mean age of $0.96 \pm 0.73$~Myr the periphery is indistinguishable in age from the Wd2 cluster, implying that star formation in the surrounding cloud set in at roughly the same time. It hosts at least 72 H$\alpha$ excess stars.

\section{The mass-accreting PMS stars}
\label{sec:Halpha}

Mass accretion onto PMS stars produces distinctive photometric and spectroscopic features. In the past, PMS stars were photometrically identified using their locus at redder colors than the MS in CMDs \citep[e.g.,][]{Hunter_95,Brandner_01,Nota_06}. A possible disadvantage of this method is the difficulty to distinguish between bona-fide PMS stars and objects that occupy the same region in the CMD (such as reddened background giants). \citet{deMarchi_10} presented a method that uses two broadband filters ($V$ and $I$ in their study) to determine the continuum emission in combination with the narrow-band H$\alpha$ filter to identify PMS stars with disk accretion. This method had been pioneered for the study of H$\alpha$ emission-line stars in young clusters by \citet{Grebel_92} and has since been widely used in multiple studies of different regions within the MW and the Magellanic Clouds \citep{deMarchi_10,deMarchi_11a,deMarchi_11b,deMarchi_13a,Beccari_10,Beccari_15,Spezzi_12}. A summary can also be found in \citet{deMarchi_15}.

The $F658N$ filter is located between and does not overlap with the $F555W$ and $F814W$ filters. To get a better characterization of the continuum contribution at the H$\alpha$ line, we thus combined the $F555W$ and $F814W$ filters to construct an interpolated $R$ filter with the following relation:

\begin{equation}
\label{eq:R-band}
R=0.237 \cdot F555W + 0.763 \cdot F814W -0.008.
\end{equation}

A detailed description is presented in Appendix~\ref{sec:R_band}.

\begin{figure}[htb]
\resizebox{\hsize}{!}{\includegraphics{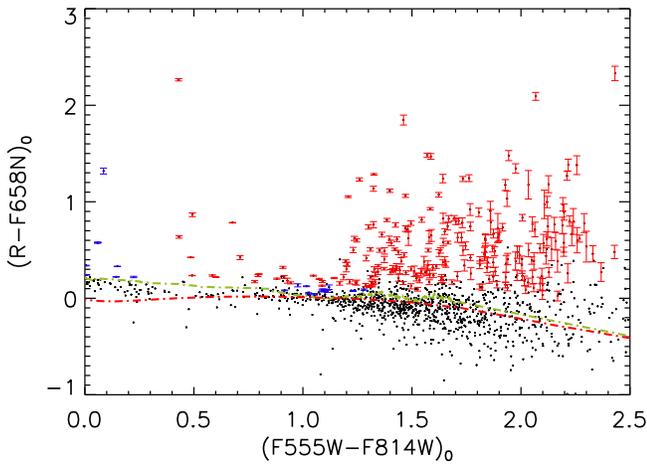}}
\caption{The $(F555W-F814W)_0$ vs. $(R-F658N)_0$ TCD of all cluster members. All stars in red fulfill the criteria of being 5 times their photometric uncertainty above the reference line of the continuum, having an $\rm{EW}(\rm{H}\alpha)>10\rm{\AA}$, and having a $(F555W-F814W)_0 > 0.2$~mag. The objects marked with blue symbols have an $\rm{EW}(\rm{H}\alpha)<10\rm{\AA}$. The red dash-dotted line represents the PARSEC 1.2S ZAMS, while the green dash-dotted line represents the photospheric continuum.}
\label{fig:Halpha_excess}
\end{figure}

The method for identifying stars with a strong H$\alpha$ emission line relies on the assumption that the majority of stars in a cluster will not have H$\alpha$ emission. We use the $(F555W-F814W)_0$ vs. $(R-F658N)_0$ TCD (see Fig.~\ref{fig:Halpha_excess}) to identify all stars with an excess emission in H$\alpha$ that is at least 5 times their photometric uncertainty above the reference line of the continuum. To do so we defined a reference template of the continuum of all stars in the given $(F555W-F814W)_0$ color range by using an average value of $(R-F658N)_0$ computed as a running mean with bin size of a 100 stars. The result is represented by the green dash-dotted line in Fig.~\ref{fig:Halpha_excess}. This method provides us with a reliable baseline because PMS stars show large variations in their H$\alpha$ excess caused by periodic mass accretion \citep[e.g.,][]{Smith_99} on an hourly or daily basis. Therefore, only a fraction of all PMS stars show H$\alpha$ excess above the continuum level at any given time.

\subsection{The H$\alpha$ excess emission}
\label{sec:Halpha_emission}

The H$\alpha$ excess emission is defined as:

\begin{equation}
\label{eq:delta_Halpha}
\Delta \rm{H}\alpha=\left(R - F658N \right)_{\rm{obs}}-\left(R - F658N \right)_{\rm{ref}};
\end{equation}
The subscript ''obs'' indicates the observed color and the subscript ''ref'' the reference template color at each $(F555W-F814W)_0$.

The combined $\Delta \rm{H}\alpha$ error is calculated as follows:

\begin{equation}
\label{eq:sigma_excess}
\sigma_{\rm{H}\alpha}=\sqrt{\left(\sigma_{F814W}^2 + \sigma_{F555W}^2 + \sigma_{F658N}^2 + \sigma_{red}^2\right)/4},
\end{equation}
where $\sigma_{F555W}$, $\sigma_{F814W}$, and $\sigma_{F658N}$ represent the photometric uncertainties of the corresponding filters and $\sigma_{red}$ is the uncertainty from the reddening map (see Paper I).

After the determination of $\Delta \rm{H} \alpha$ it is straightforward to calculate the H$\alpha$ luminosity $L(H\alpha)$:

\begin{equation}
\label{eq:Halpha_luminosity}
L(\rm{H}\alpha)= 10^{\frac{\Delta \rm{H}\alpha}{-2.5}} \cdot \rm{PHOTFLAM}_{F658N} \cdot \lambda_{\rm{P}}^{F658N} \cdot 4\pi d^2.
\end{equation}

Here \texttt{PHOTFLAM} is the inverse sensitivity of the instrument and has a value of $\rm{PHOTFLAM}_{F658N}=1.98\cdot 10^{-18}$~ergs~cm$^{-2}$~s$^{-1}$~\AA$^{-1}$. $\lambda_{\rm{P}}^{F658N}$ is the pivot wavelength of the $F658N$ filter with a value of $6583.9~\rm{\AA}$. $d=4.16$~kpc is the distance of Wd2.

\begin{figure}[htb]
\resizebox{\hsize}{!}{\includegraphics{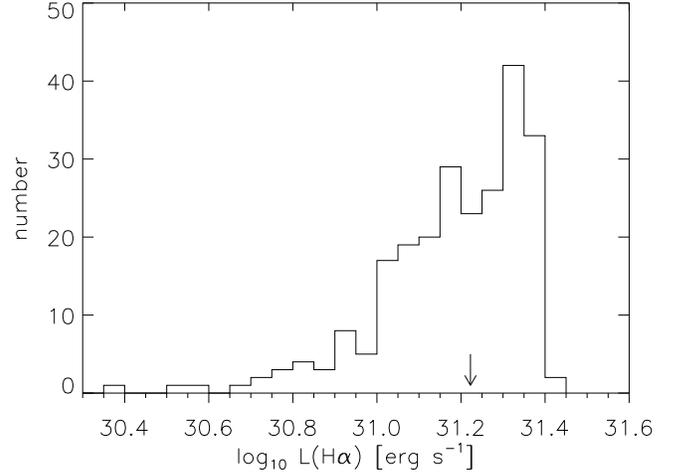}}
\caption{The histogram of the H$\alpha$ emission flux distribution of our 240 bona-fide PMS stars 5-$\sigma$ above the continuum and $(F555W-F814W)_0>0.2$~mag. The arrow marks the median value of $(4 \pm 0.36) \cdot10^{-3} L_\odot$. The bin size is 0.05~dex.}
\label{fig:Halpha_luminosity_distribution}
\end{figure}

In Fig.~\ref{fig:Halpha_luminosity_distribution} we show the distribution of the H$\alpha$ luminosity. The median H$\alpha$ luminosity is $L(\rm{H}\alpha) = 1.67\cdot10^{31}$~ergs~s$^{-1} = (4 \pm 0.36) \cdot10^{-3} L_\odot$ with a total number of 240 H$\alpha$ excess emitting stars. Additionally, we excluded all objects with $(F555W-F814W)_0<0.2$~mag for being possible Ae/Be stars \citep[e.g.,][]{Scholz_07}.

At this point we should note that the ACS $F658N$ filter is broader than a typical H$\alpha$ filter so a small portion of the \ion{N}{2} doublet at $6549\rm{\AA}$ and $6585\rm{\AA}$ falls into the H$\alpha$ filter (see Fig.~13 in Paper I). Using synthetic spectral lines from the \ion{H}{2} Regions Library \citep{Panuzzo_03} and convolving their strength with the throughput curve of the $F658N$ filter, calculated with the \texttt{bandpar} module of Synphot \footnote{Synphot is a product of the Space Telescope Science Institute, which is operated by AURA for NASA.}\citep{Synphot}, we get a contribution of 0.59\% and 3.1\% to the flux of the H$\alpha$ line. This contamination is a systematic effect and affects all stars in the same way. The combined photometric uncertainty, including the one of the color excess map used to deredden our photometry (Paper I), adds up to 8.2\% for $L(\rm{H}\alpha)$ and dominates the uncertainty. The uncertainty of 0.33~kpc in the distance of Wd2 \citep{Vargas_Alvarez_13,Zeidler_15} leads to an overall uncertainty of $L(\rm{H}\alpha)$ of $\sim 15\%$. 

\subsection{The equivalent width}
\label{sec:equvalent_width}

We use the EW of the H$\alpha$ line to separate PMS stars from those whose H$\alpha$ excess is due to chromospheric activity \citep[equivalent width, EW $\approx 3 \rm{\AA}$;][and references therein]{Panagia_00}. Because of the small photometric errors for bright stars, the 5$\sigma$ threshold is not sufficient to obtain a PMS sample that lies well above the continuum emission. \citet[][and references therein]{Panagia_00} showed that using an EW$>10\rm{\AA}$ is sufficient as an additional selection criterion to select stars well above the continuum. 

The EW gives a well-defined, comparable measurement of the strength of a line above the continuum. It is defined as:
\begin{equation}
\label{eq:def_EW}
W_{eq}=\int\left(1-P_{\lambda}\right)d\lambda,
\end{equation} 
with $P_{\lambda}$ being the line profile. In the following we always consider the absolute value in comparison of $W_{\rm{eq}}$\footnote{One should keep in mind that while looking at emission lines their EW is by definition negative.}. In the case of H$\alpha$ falling completely inside the filter width, eq.~\ref{eq:def_EW} can be calculated with the following relation:

\begin{equation}
\label{eq:EW}
W_{eq}(\rm{H}\alpha)=\rm{RW} \cdot \left( 1-10^{-0.4 \cdot (\rm{H}\alpha - \rm{H}\alpha^{cont})} \right),
\end{equation}

where RW=$74.96\rm{\AA}$ represents the rectangular width of the filter obtained with Synphot. H$\alpha$ is the observed H$\alpha$ magnitude while H$\alpha^{cont}$ is the pure H$\alpha$ continuum. This was determined using the $F555W_0$ and $F814W_0$ magnitudes of the same objects with $m_{\rm{H}\alpha^{cont}}= 0.381m_{F555W}+0.0619m_{F814W}-0.156$ \citep[determined with Synphot, see Appendix of][]{deMarchi_10}. \citet{deMarchi_10} also showed that this transformation does not significantly change with metallicity.

\begin{figure}[htb]
\resizebox{\hsize}{!}{\includegraphics{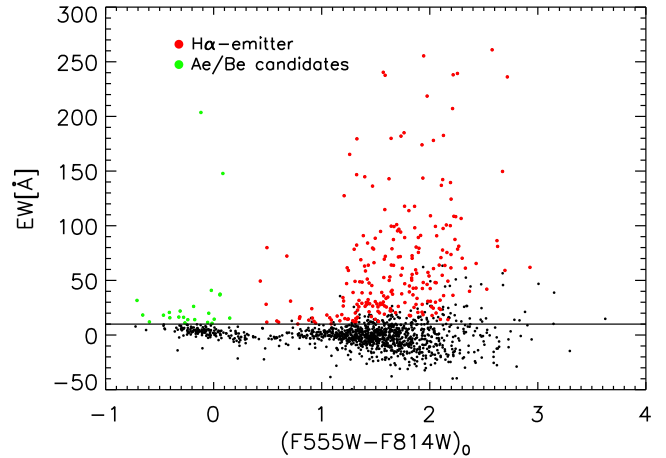}}
\caption{The distribution of the EWs for all cluster members. On the x-axis the $(F555W-F814W)_0$ color is plotted and on the y-axis the EW in \AA ngstrom. Objects in red represent the 240 bona fide H$\alpha$ emission PMS stars that fulfill the selection criteria. The Ae/Be candidates are represented as green dots.}
\label{fig:EW}
\end{figure}

We find that 74.6\% of all H$\alpha$ excess sources have an EW$>10\rm{\AA}$. Additionally removing the 24 Ae/Be candidates (red dots in Fig.~\ref{fig:F814W-F160W_F814W}) leaves us with 240 objects (67.7\%). In Fig.~\ref{fig:EW} we show the EW distribution, including the 240 stars considered to be H$\alpha$-emitting PMS stars (red dots) and the 24 Ae/Be candidates (green dots). The locus of the H$\alpha$ excess stars in the $F814W_0$ vs. $(F814W-F160W)_0$ CMD is shown in Fig.~\ref{fig:F814W-F160W_F814W}. The majority of the Ae/Be candidates lies, as expected, in the MS and TO regime (blue shaded area in Fig.~\ref{fig:F814W-F160W_F814W}).

\subsection{The Ae/Be star candidates}
\label{sec:AeBe_candiates}
In Sect.~\ref{sec:equvalent_width} we classified all stars showing an H$\alpha$ excess 5$\sigma$ above the continuum with $(F555W-F814W)_0<0.2$~mag as possible Ae/Be stars \citep[e.g.,][]{Scholz_07}. This led to a number of 24 Ae/Be candidates. Some of these stars are located in the same area of the $F814W_0$ vs. $(F814W-F160W)_0$ CMD as the PMS (see Fig.~\ref{fig:F814W-F160W_F814W}). Classical Ae/Be stars can show IR emission caused by gaseous circumstellar disks \citep[e.g.,][]{Hillenbrand_93} which can lead to a higher $(F814W-F160W)$ color. To check whether our candidates are classical Ae/Be stars or Herbig Ae/Be stars we also analyzed their locus in the $F814W_0$ vs. $(F555W-F814W)_0$ CMD. As can be seen all 24 stars are located well outside the area covered by the PMS. \citet{Subramaniam_06} compared the spectra of classical Ae/Be stars and Herbig Ae/Be stars and showed that the slope of the continuum flux widely differs leading to blue $(F555W-F814W)$ colors for classical Ae/Be stars and red $(F555W-F814W)$ colors for Herbig Ae/Be stars. Since, by our selection criterion, all of our candidates have colors bluer than $(F555W-F814W)_0 < 0.2$~mag we can state that these stars are all Ae/Be candidates.

\subsection{Ages and masses of the PMS stars}
\label{sec:stellar_properties}

To determine the mass accretion rates onto the PMS stars, it is necessary to know the properties of the central stars, such as the effective temperature, mass, luminosity, and age. We estimated these stellar properties from the PARSEC 1.2S evolutionary models \citep{Bressan_12}. We determined the stellar parameters, as well as their ages, from the isochrones closest to each individual star for a grid of five isochrones (0.1~Myr, 0.25~Myr, 0.5~Myr, 1.0~Myr, and 2.0~Myr). In Paper I, we assumed a solar metallicity of $Z_\odot=0.019$, based on the hypothesis that, as a member of the thin disk, Wd2 would have solar abundance. The isochrones used in Paper I for $Z_\odot=0.019$ did not reproduce the slope of the PMS evolutionary phase very well. For the latest PARSEC 1.2S models, \citet{Bressan_12} used a different metallicity $Z_\odot=0.0152$ for the Sun. They used the element abundances compiled by \citet{Grevesse_98} and adopted revised values from \citet[][and references therein]{Caffau_11}. In the PMS region, these new isochrones have a steeper slope and, therefore, reproduce better the colors of our data. Throughout this paper we use this revised Solar metallicity.

\begin{figure}[htb]
	\resizebox{\hsize}{!}{\includegraphics{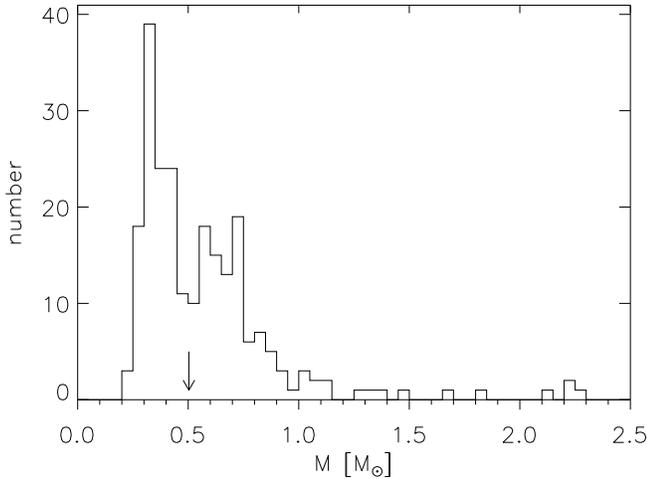}}
 	\caption{The determined mass distribution of the 240 bona-fide mass-accreting PMS stars is shown. Most of the stars are in the sub-solar regime peaking at $\sim 0.25~M_\odot$. The arrow marks the mean mass $M=0.50~M_\odot$.}
 	\label{fig:mass_distribution}
\end{figure}

The stellar evolution tables of the PARSEC 1.2S models list the effective (photospheric) temperature ($T_{\rm{eff}}$), the mass ($M_\star$), and the bolometric luminosity ($L_\star$) of each star. In Fig.~\ref{fig:mass_distribution} we show the mass distribution of the 240 bona-fide mass-accreting PMS stars. The vast majority of the stars has sub-solar mass.
 
\section{Accretion luminosity and mass accretion rate}
\label{sec:acc_L_and_M}
The source of the bolometric accretion luminosity ($L_{acc}$) is radiation emitted by the accretion process of the disk onto its central star \citep{Hartmann_98}. This leads to a connection between the H$\alpha$-excess luminosity $L(\rm{H}\alpha)$, produced by the same process, and the accretion luminosity. For the logarithmic values of $L(\rm{H}\alpha)$ and $L_{\rm{acc}}$, theoretical models of \citet{Muzerolle_98b} predict a slope of unity for low accretion rates and shallower slopes for higher accretion rates. The empirical fit of $\log_{10}L_{acc}$ vs. $\log_{10}L(\rm{H}\alpha)$ for 14 members of IC~348 in the Taurus-Auriga association by \citet{Dahm_08} is characterized by a slope of $1.18\pm0.26$. Taking into account the larger uncertainty associated with our data and the fact that we most likely have a sample with a variety of accretion rates, we will use eq.~5 of \citet{deMarchi_10} obtained from the data presented in \citet{Dahm_08}. On this basis $L_{\rm{acc}}$ is connected with $L(\rm{H}\alpha)$ the following way:

\begin{equation}
\label{eq:Lacc_LHalpha}
\log_{10}(L_{\rm{acc}})=\left(1.72 \pm 0.47 \right)  + \log_{10}L(\rm{H}\alpha).
\end{equation}

\begin{figure}[htb]
	\resizebox{\hsize}{!}{\includegraphics{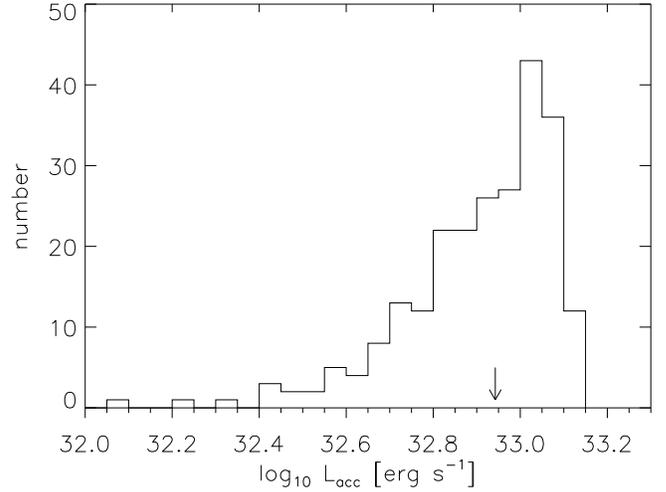}}
	\caption{The accretion luminosity distribution of our 240 bona-fide PMS stars. The arrow marks the median value of $8.76 \cdot10^{32}~\rm{erg}~\rm{s}^{-1}$. The bin size is 0.05~dex.}
	\label{fig:Lacc_luminosity_distribution}
\end{figure}

The uncertainty of $\pm0.47$ shows how difficult it is to find a relation between the two observables, yet it is the best relation that we can use to relate $L(\rm{H}\alpha)$ to $L_{\rm{acc}}$. Applying the transformation to the accretion luminosity for our objects gives us a median value $L_{\rm{acc}}=0.23 \pm 0.029~L_\odot$. The errors represent only the photometric uncertainties. The accretion luminosity distribution is shown in Fig.~\ref{fig:Lacc_luminosity_distribution}.

We can now use the free-fall equation to link the accretion luminosity to the accretion mass rate $\dot M_{acc}$ in the following manner:

\begin{equation}
\label{eq:Macc}
L_{acc}=\frac{G M_\star \dot M_{acc}}{R_\star} \left( 1- \frac{R_\star}{R_{in}} \right).
\end{equation}

$G$ is the gravitational constant, $M_\star$ and $R_\star$ are the stellar mass and radius, while $R_{in}$ is the inner radius of the accretion disk. $T_{\rm{eff}}$ can be used together with the bolometric luminosity to calculate the stellar radius $R_\star$. Following \citet{Gullbring_98}, we assume $R_{in} \approx 5 R_\star$ for all objects. Combining now eq. \ref{eq:Lacc_LHalpha} and eq. \ref{eq:Macc}, we get the mass accretion rate as a function of $L(\rm{H}\alpha)$:
 
\begin{eqnarray}
\label{eq:Macc_LHalpha}
\log_{10}\frac{\dot M_{acc}}{M_\odot \rm{yr}^{-1}}=\nonumber \\
= -7.39 + \log_{10}\frac{L_{acc}}{L_\odot} + \log_{10} \frac{R_\star}{R_\odot} - \log_{10} \frac{M_\star}{M_\odot}\nonumber \\
= \left(-5.67 \pm 0.47 \right) + \nonumber \\
+ \log_{10}\frac{L(\rm{H}\alpha)}{L_\odot} + \log_{10} \frac{R_\star}{R_\odot} - \log_{10} \frac{M_\star}{M_\odot}.
\end{eqnarray}

Calculating the mass accretion rates for our 240 H$\alpha$ excess sources gives a median mass accretion rate $\dot M = 4.43 \cdot 10^{-8} M_\odot \rm{yr}^{-1}$.

The error on the mass accretion rate associated with the uncertainties in the photometry is $0.363 \cdot 10^{-8} M_\odot \rm{yr}^{-1}$. Another error source is the determination of the stellar parameters $L_\star$, $M_\star$, and $T_\star$. To examine these errors we varied each of the stellar parameters by $\pm 1\%$, $\pm 5\%$, and $\pm 10\%$. This results in an uncertainty on the mass accretion rate of $\pm2.7\%$, $\pm10.9\%$, and $\pm18.7\%$, respectively.

The accretion luminosity can contribute up to 30\% to the bolometric luminosity of a mass-accreting PMS star, with a median contribution of 15\%. To determine the stellar properties, we used the CMD based on the $F814W$ and $F160W$ filters (see Sect.~\ref{sec:stellar_properties}). The $F658N$ filter (H$\alpha$) does not overlap in wavelength with any of the broad-band filters used in this study (see Fig.~13 in Paper~I). Therefore, we can say that the contribution of the accretion luminosity to the bolometric luminosity is not influencing our results and conclusions since we are not using bolometric luminosities, but instead the luminosities in the broadband filters listed above.

\subsection{Mass accretion rate as a function of stellar age}

\begin{figure}[htb]
	\resizebox{\hsize}{!}{\includegraphics{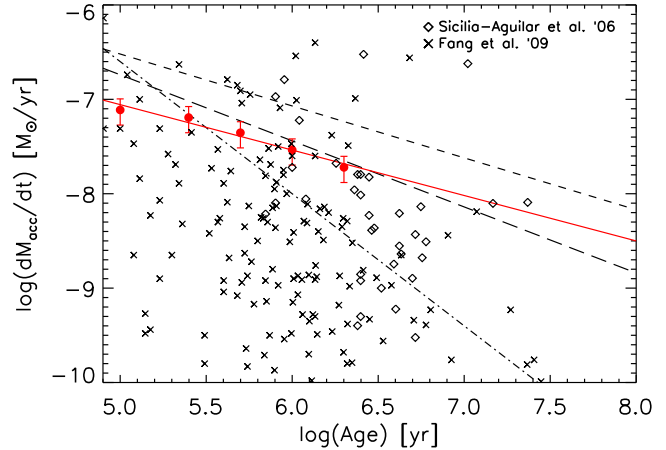}}
	\caption{The median values of the mass accretion $\dot M$ vs. the stellar age (red solid line). The error bars represent the uncertainty. The mass accretion decreases with age with a slope of $\eta=0.48 \pm 0.04$. The long-dashed and short-dashed lines represent the age-mass accretion rate relation derived by \citet{deMarchi_13a} for the two clusters NGC~602 and NGC~346, respectively. The crosses and diamonds show the mass-accreting stars in Tr~37 \citep{Sicilia-Aguilar_06} and Orion Giant Molecular Cloud \citep{Fang_09}. The dash-dotted line is the relation derived by \citet{Hartmann_98} for the viscous disk evolution.}
	\label{fig:M_acc_vs_age}
\end{figure}

\begin{figure}[htb]
	\resizebox{\hsize}{!}{\includegraphics{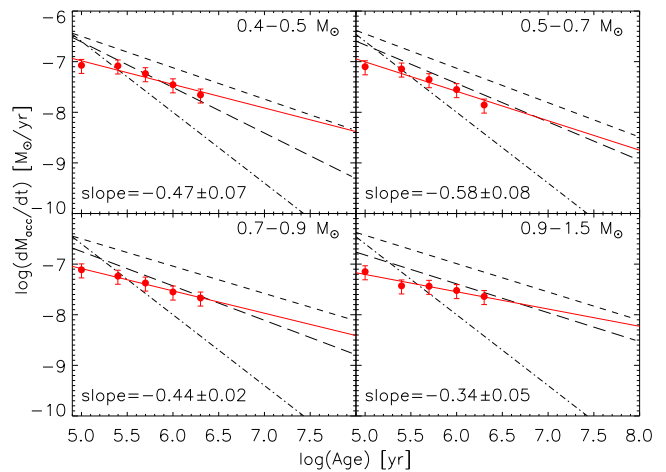}}
	\caption{The median values of the mass accretion $\dot M$ per age bin vs. the stellar age (red solid line) for different mass bins (0.4--0.5~M$_\odot$, 0.5--0.7~M$_\odot$, 0.7--0.9~M$_\odot$, 0.9--1.5~M$_\odot$). The error bars represent the uncertainty. The long-dashed and short-dashed lines represent the age-mass accretion rate relation derived by \citet{deMarchi_13a} for these mass bins for the two clusters NGC~602 and NGC~346, respectively. The dashed-dotted line is the relation derived by \citet{Hartmann_98} for viscous disk evolution.}
	\label{fig:M_acc_vs_age_massbins}
\end{figure}

We calculated the median mass accretion rate for each age bin (0.1~Myr, 0.25~Myr, 0.5~Myr, 1.0~Myr, and 2.0~Myr; red dots in Fig.~\ref{fig:M_acc_vs_age} and Fig.~\ref{fig:M_acc_vs_age_massbins}) and found that the mass accretion rate decreases with the stellar age ($\dot M \propto t^{-\eta}$, with $\eta=0.48 \pm 0.04$ indicated by the red line in Fig.~\ref{fig:M_acc_vs_age}). \citet{Hartmann_98} determined a slope of $\eta \approx 1.5 -2.7$ with large uncertainties up to $\Delta \eta = 0.7$ for the viscous disk evolution, and stated that ''this slope is poorly constrained'' (dash-dotted line in Fig.~\ref{fig:M_acc_vs_age}). We also plotted the age-mass accretion relation derived by \citet{deMarchi_13a} for the two clusters NGC~602 and NGC~346 represented by the long-dashed and short-dashed line, respectively. The slopes and mass accretion rates are similar to the ones of Wd2. Comparing the mass accretion rates estimated in this paper with the data collected by \citet{Calvet_00} from multiple sources \citep[Fig.4,][and references therein]{Calvet_00}, we can conclude that our mass accretion rates are comparable to those data.

In Fig.~\ref{fig:M_acc_vs_age_massbins} we show the decrease of the mass accretion rates with time for different mass bins (0.4--0.5~$M_\odot$, 0.5--0.7~$M_\odot$, 0.7--0.9~$M_\odot$, and 0.9--1.5~$M_\odot$). The error-weighted fit shows an overall decrease of the slope of the relation and is consistent with what \citet{deMarchi_13a} found.

\subsection{The spatial distribution of mass accreting PMS stars}
\label{sec:Halpha_pop_dist}

We showed that the mass accretion rate in Wd2 decreases with stellar age as was predicted by e.g., \citet{Hartmann_98} and \citet{Sicilia-Aguilar_06}. Another point to take into account is the high number of luminous OB stars especially in the cluster center (MC and NC). These massive, luminous stars emit a high amount of far ultra-violet (FUV) flux that can erode nearby circumstellar disks \citep[e.g.,][]{Clarke_07}. \citet{Anderson_13} studied the effects of photoevaporation of disks due to their close proximity to massive OB stars. They found that, depending on the viscosity of the disk, most disks are completely dispersed within 0.5--3.0~Myr. This timescale is so short that, if this effect was present in the center of Wd2, we should already detect this decrease. In addition to the timescale, also the distance to the FUV source plays an important role. The results of \citet{Anderson_13} for the Orion Nebula Cluster indicate that the influence of OB stars only plays a role up to a distance of 0.1--0.5~pc.

\begin{figure*}[htb]
	\resizebox{\hsize}{!}{\includegraphics{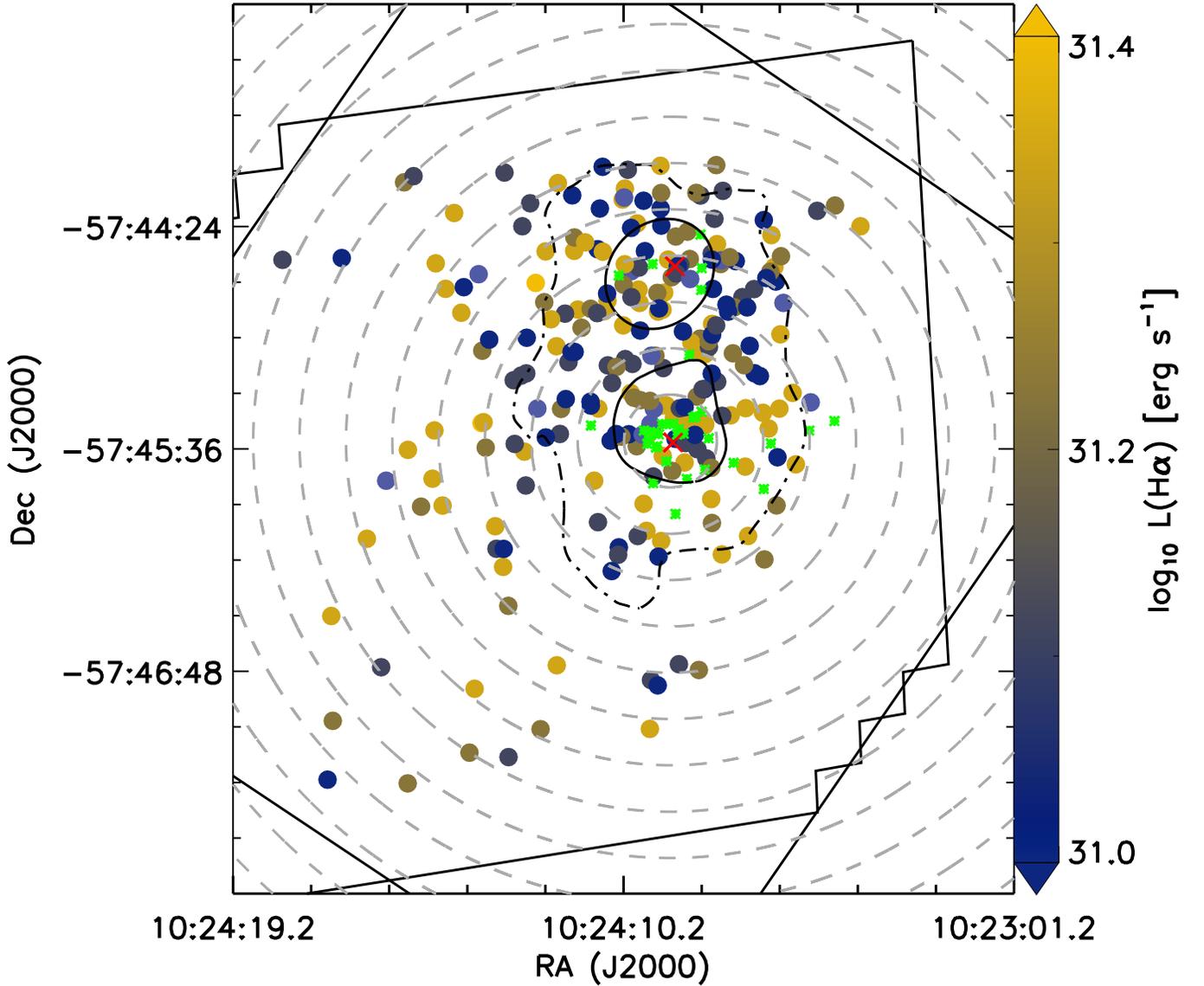}}
	\caption{The 240 H$\alpha$ excess stars are here shown color-coded with their respective H$\alpha$ excess luminosity indicated on the color bar. The two solid oval contours mark the MC and the NC while the dash dotted contour marks the overall Wd2 cluster. The gray dashed circles indicate the distance from the MC center in steps of $15''$ or 0.3~pc. The green asterisks mark all know O and B stars while the red crosses mark the geometric center for all known O and B stars within 0.5~pc of the MC and NC, respectively. For orientation purposes, the thick black straight lines mark the outline of the FOV of the survey area.}
	\label{fig:Halpha_spatial_dist}
\end{figure*}

In Wd2 we only see a 2D projection of the 3D distribution of the stars. Assuming that the MC and the NC are approximately spherical, their distribution in the z-direction does not differ from that in x and y. In Fig.~\ref{fig:Halpha_spatial_dist} the spatial locations of all 240 H$\alpha$ excess stars are plotted, color-coded with the amount of the H$\alpha$ excess luminosity. The green asterisks mark all known OB stars in RCW~49. As reference, the FOV of the survey area and the contours of the MC the NC (solid contours), and the Wd2 cluster (dashed dotted contour) are over-plotted. The gray, dashed circles indicate the radial distance of the center of the MC in steps of $15''$ or 0.3~pc. The MC is located entirely within a radius of 0.5~pc.

\begin{figure*}[htb]
	\plottwo{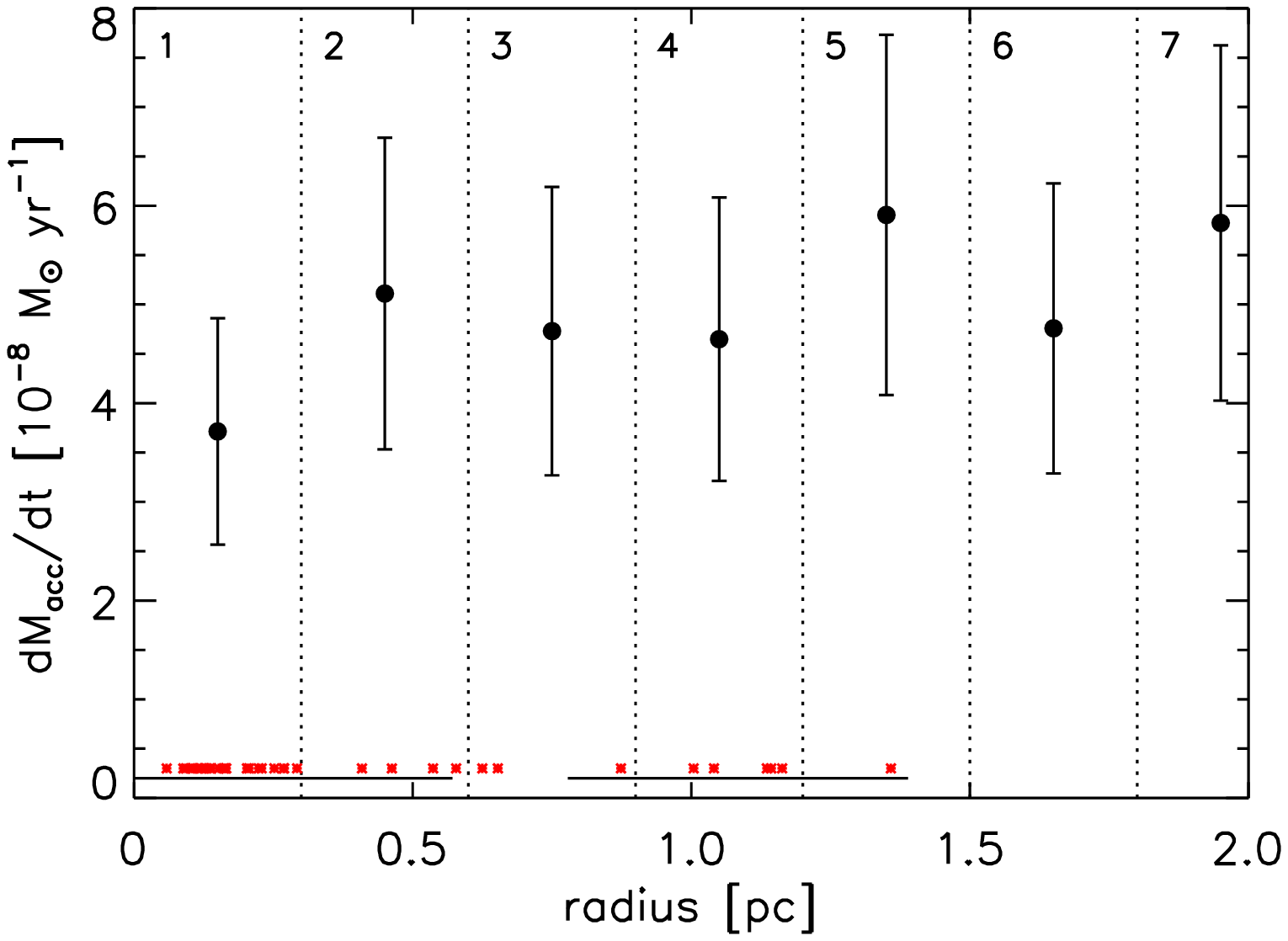}{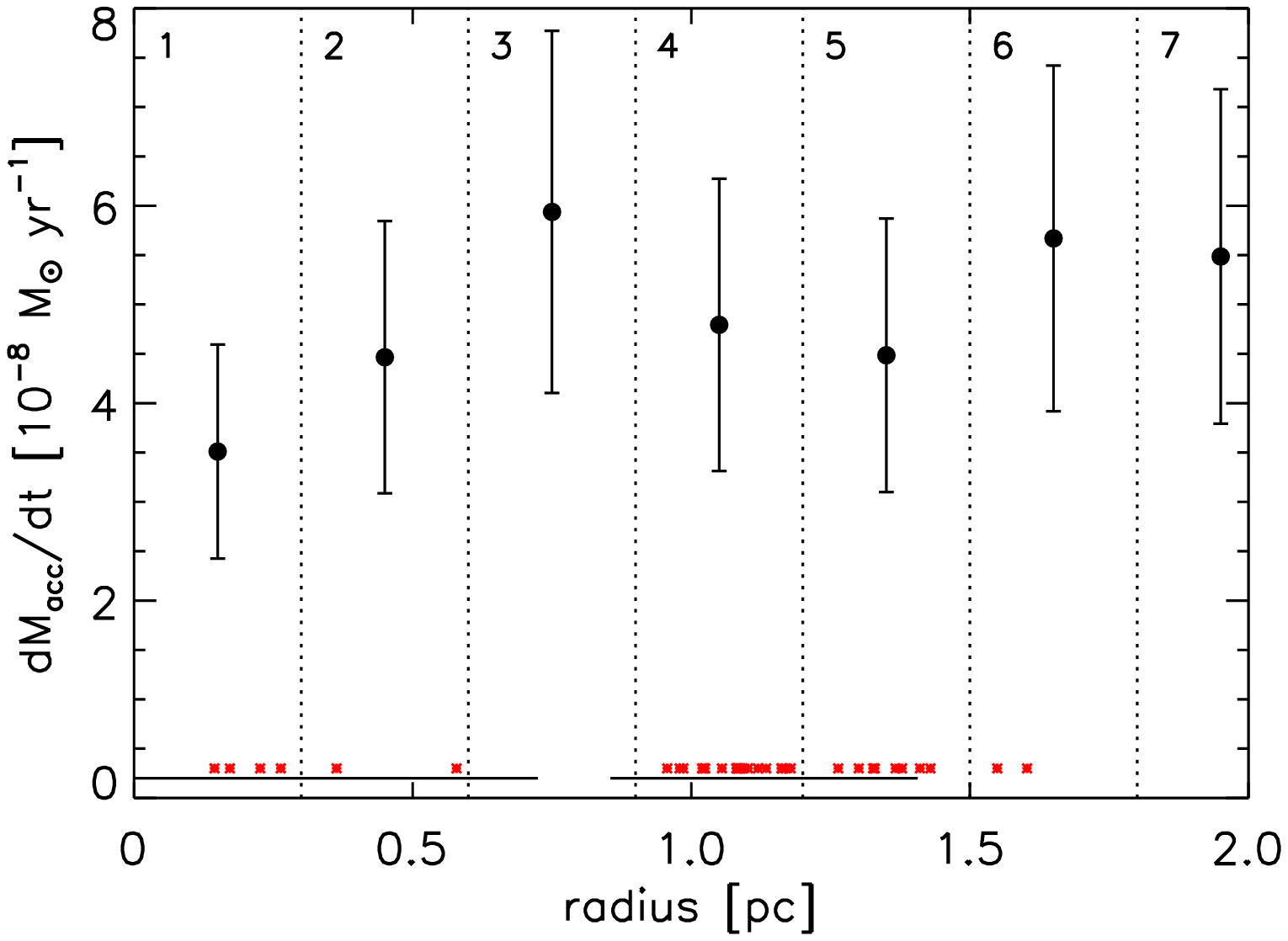}
	\caption{The mean mass accretion rate as a function of distance from the projected geometric center of the OB stars in the MC and NC going outwards in annuli of $15''$ (0.3~pc). \textbf{Left: }The mean mass accretion rate for the MC. \textbf{Right: }The mean mass accretion rate for the NC. The two bars at the bottom mark the spatial extent of the MC and NC while all OB stars are represented as red asterisks. The vertical dotted lines indicate the annuli of $15''$, numbered from the center outwards starting with 1.}
	\label{fig:radial_Halpha_dist}
\end{figure*}

The mass accretion rate is linked to the H$\alpha$ excess luminosity (including the dependence on the mass and age of the star, see eq. \ref{eq:Macc_LHalpha}). The median mass accretion rate in the Wd2 cluster outskirts is $4.84~\cdot 10^{-8}~M_\odot \rm{yr}^{-1}$. In comparison, the median mass accretion rate in the MC and NC is $3.32~\cdot 10^{-8}~M_\odot \rm{yr}^{-1}$ and $3.12~\cdot 10^{-8}~M_\odot \rm{yr}^{-1}$, respectively. The lower mass accretion rate in the MC and NC is caused by the presence of a high number of OB stars in their centers. To further analyze this we calculated the projected geometric centers of all OB stars within 0.5~pc of each of the peak positions of the MC and NC. These peak positions are represented with red crosses in Fig.~\ref{fig:Halpha_spatial_dist}. The geometric center of the OB stars in the MC almost coincides with the MC peak position ($d=0.98''$). For the NC the geometric center of all OB stars within 0.5~pc from the NC peak position is $5.73''$. We used these centers to calculate the mean mass accretion rate per annulus going outwards in steps of $15''$ or 0.3~pc. The results for both the MC and the NC are represented in Fig.~\ref{fig:radial_Halpha_dist}. Each annulus was given a number for an easier reference in the text, starting with 1 in the center (see Fig.~\ref{fig:Halpha_spatial_dist} and \ref{fig:radial_Halpha_dist}).

With $3.71~\cdot 10^{-8}~M_\odot \rm{yr}^{-1}$ for the MC and $3.51~\cdot 10^{-8}~M_\odot \rm{yr}^{-1}$ for the NC the mass accretion rate for both clumps is the lowest in their respective OB-star-defined center.

Using the MC center as origin for the radial analysis results in an increase of $\dot M$ by $\sim60\%$ to $5.11~\cdot 10^{-8}~M_\odot \rm{yr}^{-1}$ within the inner $30''$ (0.6~pc), going from the first to the second annulus. The first annulus (innermost $15''$ or 0.3~pc) includes 23 of the OB stars, while the second annulus ($15''$--$30''$ or 0.3--0.6~pc) includes only 4 OB stars. The larger distance to the OB stars of the second annulus explains the steep increase of the mass accretion rate. Going further outwards to the annuli 3 and 4 ($30''$--$60''$ or 0.6--1.2~pc), $\dot M$ decreases by $\sim 9\%$ to $4.72~\cdot 10^{-8}~M_\odot \rm{yr}^{-1}$. These two annuli contain 9 OB stars, while 4 OB stars are in the NC. These stars probably cause the decrease of the mass accretion rate. In annulus 5 ($60''$--$75''$ or 1.2--1.5~pc) the mass accretion rate rises to $5.90~\cdot 10^{-8}~M_\odot \rm{yr}^{-1}$. From this point outwards, the OB stars do not affect the mass accretion rate anymore and the fluctuations in $\dot M$ are caused by small-number statistics of the H$\alpha$ excess stars ($\le 10$). Overall we can see a trend of an increase of the mass accretion rate with increasing distance from the OB stars, indicating that the PMS accretion disks are being rapidly destroyed by the far-ultra-violet radiation emitted by the OB population.

Using the NC center as origin for the radial analysis gives a similar result as for the MC. Going outwards from the center the mass accretion rate increases to $4.47~\cdot 10^{-8}~M_\odot \rm{yr}^{-1}$ in the second annulus and to $5.94~\cdot 10^{-8}~M_\odot \rm{yr}^{-1}$ in the third annulus. This corresponds to an increase of $\sim 68\%$ within the inner $45''$ (0.9~pc). While we have 6 OB stars in the innermost annuli, the number drops to 0 in the third. PMS stars in the third annulus are located in between the two clumps, $\gtrapprox 0.5$~pc away from the OB stars, and therefore out of their sphere of influence of 0.1--0.5~pc \citep{Clarke_07}. The annuli 4 and 5 ($45''$--$75''$ or 0.9--1.5~pc) cover the area of the MC with a total number of 28 known OB stars. Their FUV radiation makes the mass accretion rate drop by $\sim20\%$ to $4.79~\cdot 10^{-8}~M_\odot \rm{yr}^{-1}$. From the next annulus outwards ($d\ge 75''$, $\ge 1.5$~pc) the mass accretion rate increases to $5.67~\cdot 10^{-8}~M_\odot \rm{yr}^{-1}$. We can thus see the same trend as when are using the MC as center of origin. With increasing distance from the OB star population, the mass accretion rate increases.

For a better overview, we summarized in Tab.~\ref{tab:mass_acc_rate} the values for the mass accretion rates for each of the annuli for the respective center of origin (MC and NC).

\floattable

\begin{deluxetable}{cccc}
	\tablecaption{The mean mass-accretion rates\label{tab:mass_acc_rate}}
	\tablehead{
		\multicolumn{1}{c}{\#} &\multicolumn{1}{c}{distance [pc]} &\multicolumn{2}{c}{$\dot M~[10^{-8}~M_\odot \rm{yr}^{-1}]$} \\
		\multicolumn{1}{c}{ } &  \multicolumn{1}{c}{ } & \multicolumn{1}{c}{MC} & \multicolumn{1}{c}{NC}
		}
	\startdata
	1			  &	0.0--0.3 & 3.71  & 3.51  \\
	2		 	  &	0.3--0.6 & 5.11  & 4.47  \\
	3		 	  &	0.6--0.9 & 4.72  & 5.94  \\
	4			  &	0.9--1.2 & 4.64  & 4.79  \\
	5 			  &	1.2--1.5 & 5.90  & 4.49  \\	
	6 			  &	1.5--1.8 & 4.76  & 5.67  \\
	7			  &	1.8--2.1 & 5.83  & 5.49  \\
	\enddata
	\tablecomments{The mean mass accretion rates as a function of distance from the projected geometric center of the OB stars in the MC and NC going outwards in annuli of $15''$ (0.3~pc). Column~1 gives the number for each annulus as used in Fig.~\ref{fig:radial_Halpha_dist}. Column~2 lists the distance of each annulus from the respective centers. Column~3 and 4 give the mean mass accretion rates in each annulus for the MC and NC, respectively.}
\end{deluxetable}

Despite a few objects in Fig.~\ref{fig:Halpha_spatial_dist} showing high H$\alpha$ excess luminosity, which may appear to lie close to OB stars due to the projection of a 3D stellar distribution onto a 2D map, the evolution of the mass accretion rate with distance to the population of luminous OB stars is consistent with theoretical studies \citep[e.g.,][]{Clarke_07} and with the observations made by \citet{Anderson_13} for the Orion Nebula Cluster and \citet{deMarchi_10} in the field around the \object{SN~1987A}. With $4.43~\cdot 10^{-8}~M_\odot \rm{yr}^{-1}$ the median mass accretion rate of the PMS stars of Wd2 is 1.5 times higher than in the region surrounding \object{SN~1987A} \citep{deMarchi_10}. Taking into account the uncertainties (see Sect.~\ref{sec:uncertainties}) and the younger age of Wd2 the mass accretion rates are in good agreement with these results (see Sect.~\ref{sec:summary} for a detailed discussion).

\section{Uncertainties in the H$\alpha$ luminosity and mass accretion rate}
\label{sec:uncertainties}
To derive and quantify the mass accretion rate and the H$\alpha$ luminosity we compared the observations of our multi-band survey (see Sect.~\ref{sec:catalog}) with theoretical models of mass accretion onto T-Tauri stars (see Sect.~\ref{sec:acc_L_and_M}) in combination with empirically derived relations (see Sect.~\ref{sec:Halpha_emission} and Sect.~\ref{sec:acc_L_and_M}). The resulting mass accretion rates are affected by different kinds of uncertainties:

\begin{itemize}
	\item The photometric uncertainties (see Paper I)
	\item The uncertainties of the extinction map, used to deredden the photometry (see Paper I)
	\item Uncertainties in the stellar evolution models
	\item Uncertainty in the adopted stellar abundance
	\item Uncertainties occurring while fitting the models to the data	
\end{itemize}

Some of these error sources were already briefly discussed in the previous sections. In the following we want to summarize and give an overview of all sources of uncertainties.

\subsection{Observational uncertainties}
\label{sec:phot_uncertainties}
In Sect. \ref{sec:catalog} we gave a brief overview of the photometric catalog fully described in Paper I. To obtain the H$\alpha$ excess emission and the mass accretion rate we used the individually dereddened photometry in the three filters $F555W$, $F814W$, and $F658N$ (see eq. \ref{eq:delta_Halpha}). This was achieved using the $E(B-V)$ gas extinction map (see Paper I). The observational uncertainties include the combined photometric uncertainties plus the uncertainty originating from the gas extinction map (see eq.~\ref{eq:delta_Halpha}). This gives a total uncertainty of 8.2\% for $L(H\alpha)$ or $0.140 \cdot 10^{-8} M_\odot \rm{yr}^{-1}$ for the mass accretion rate.

Additionally, the \ion{N}{2} doublet partially falls into the $F658N$ filter width causing a possible overestimation of the H$\alpha$ flux by 0.59\% and 3.1\% (see Sect. \ref{sec:Halpha_emission}).

\subsection{The locus of the isochrones}

The loci of the stars relative to the PARSEC 1.2S isochrones \citep{Bressan_12} in the $F814W_0$ vs. $(F814W-F160W)_0$ CMD play an important role in defining the stellar and cluster properties.

\begin{itemize}
	\item[1)] The heliocentric distance was determined using the TO region in combination with the results of the spectroscopic analysis of \citet{Vargas_Alvarez_13}. The uncertainty in the distance $d=4.16$~kpc between Wd2 and the Sun is $\Delta d=0.33$~kpc (see Paper I). This gives a $L(H\alpha)$ uncertainty of 15\% (see eq.~\ref{eq:Halpha_luminosity}).
	\item[2)] The loci of the stars in the CMD define the stellar properties, such as masses, temperatures, bolometric luminosities, and stellar radii. To estimate the possible uncertainties we varied each of the stellar parameters by $\pm1\%$, $\pm5\%$, and $\pm10\%$. This gives overall uncertainties in the mass accretion rate of $\pm2.7\%$, $\pm10.9\%$, and $\pm18.7\%$, respectively.
\end{itemize}

\subsection{The stellar metallicity}

Based on the hypothesis that Wd2 is a member of the thin disk, we assumed Solar metallicity \citep[$Z=0.0152$,][]{Caffau_11}. Nevertheless, since we cannot determine the true metallicity of the cluster, we estimate the effects on the mass accretion rate by modifying the metallicity of the stellar evolution models. We varied the assumed metallicity of $Z=0.0152$ by $\pm 10\%$ (to $Z=0.0137$ and $Z=0.0162$) and $\pm 25\%$ (to $Z=0.0114$ and $Z=0.019$). Increasing the metallicity by $10\%$ and $25\%$ decreases the mass accretion rate by $4\%$ and $7\%$, while the decrease of the metallicity by $10\%$ and $25\%$ leads to an increase of the mass accretion rate by $2\%$ and $7\%$, respectively. Considering the small dependence of the mass accretion rate on metallicity and the fact that the distribution of stars in Wd2 in our CMDs (see Fig.~\ref{fig:F814W-F160W_F814W} and Fig.~\ref{fig:F555W-F814W_F814W}) is best represented by isochrone models of Solar metallicity, supports our assumption of Solar metallicity.

\subsection{Geometrical alignment}
\label{sec:stat_uncertainties}

The geometrical orientation of the disks plays an important role for the emitted light that we can detect. Here, we are referring especially to the inclination of a disk relative to the sky plane. Two major cases need to be distinguished:

\begin{itemize}
	\item[1)] A large enough inclination, meaning the orientation is almost edge-on, leads to an obscuration of the star by its surrounding disk. The flux at short (UV, optical, and NIR) wavelengths is blocked by the disk material. Therefore, these objects are not detected in our optical/NIR catalog. Assuming a flared-disk model and comparing with the spectral-energy distribution (SED) modeling of \citet{Chiang_99} this happens at inclination angles $i \gtrsim 55^\circ$. We can conclude that we miss $\sim 40\%$ of the H$\alpha$ excess stars due to this geometrical effect. 
	\item[2)] A moderately small inclination ($i \lesssim 55^\circ$, face-on), so the disk does not block the light emitted by the host star. The inclination should play a major role for the shape of the emission lines \citep[e.g.,][]{Muzerolle_01,Kurosawa_06,Kurosawa_12} since stellar rotation broadens the emission lines. Comparing this effect with observations \citep{Appenzeller_13} no significant result has been found yet. Most likely this is because of the the small sample of stars studied so far. \citet[][observations]{Appenzeller_05} and \citet[][theory]{Kurosawa_06} found a dependence of the EW on the inclination angle. Using a larger sample of stars, \citet{Appenzeller_13} could not find this specific correlation. This leads to the conclusion that, even if there is an effect due to the rotation of the disk of the PMS stars, at the moment there is no way of further quantifying it. The effect on line broadening due to stellar rotation does not play a role for our photometric observations because the broadening is less than the filter width and so the original flux is fully detected.
\end{itemize}

Because we can only detect disk-accreting PMS stars via the H$\alpha$ excess if the disk is not blocking the light of its central star \citep{Chiang_99}, the effects on the colors and luminosities of the PMS stars caused by disk obscuration, and a resulting uncertainty in age, are minor.

We should note that some of the ionizing energy may possibly escape without having an effect on the local surrounding gas. This causes an underestimation of the mass accretion rate. This is also the case for all other studies based on hydrogen emission.

Altogether, the different sources of uncertainties are presented in Tab.~\ref{tab:uncertainties}. They add up to a total uncertainty in the H$\alpha$ luminosity $L(\rm{H}\alpha)=\left(1.67 \pm 0.449\right) \cdot 10^{-31}~\rm{erg}~\rm{s}^{-1}$ (26.9\%). The total uncertainty on the mass accretion rate (assuming that the stellar parameters are known to a precision of $<5\%$) amounts to $\Delta \dot M_{\rm{acc}}=1.768 \cdot 10^{-8}~M_\odot~\rm{yr}^{-1}$ (39.9\%).

\floattable
\begin{deluxetable}{lrrr}
	\tablecaption{The sources of uncertainties \label{tab:uncertainties}}
	\tablehead{
		\multicolumn{1}{c}{Source} 	&\multicolumn{1}{c}{Uncert.} & \multicolumn{1}{c}{$\sigma(\dot M_{acc})$} & \multicolumn{1}{c}{$\sigma(L(\rm{H}\alpha))$}\\
		\multicolumn{1}{c}{}		&\multicolumn{1}{c}{[\%]} & \multicolumn{1}{c}{$[10^{-8}~M_\odot~\rm{yr}^{-1}]$} & \multicolumn{1}{c}{$[10^{-31}~\rm{erg}~\rm{s}^{-1}]$}
	}
	\startdata
	Photometry			& 8.2	 & 0.363 & 0.136 \\
	\ion{N}{2}-doublet	& 3.7	 & 0.164 & 0.062 \\
	Dist. modulus		& 15	 & 0.665 & 0.251 \\
	Stellar models		& 11	 & 0.487 & ---	 \\
	Metallicity			& 2		 & 0.089 & ---	 \\ [0.1cm]
	Total			& 39.9/26.9	 & 1.768 & 0.449 \\
	\enddata
	\tablecomments{The summary of the different sources of uncertainty. The two values of the total uncertainty percentage correspond to the mass accretion rate and to the H$\alpha$ luminosity, respectively.}
\end{deluxetable}

\section{Summary and Conclusions}
\label{sec:summary}
In this paper we examined the PMS population of RCW~49 using our recent optical and near-infrared HST dataset of Wd2, obtained in 6 filters ($F555W$, $F658N$, $F814W$, $F125W$, $F128N$, and $F160W$; for more details see Paper I).

To analyze the PMS population of Wd2 we determined the stellar parameters ($T_{\rm{eff}}$, $L_{\rm{bol}}$, and $M_\star$) using the PARSEC 1.2S \citep{Bressan_12} stellar evolution models. We estimated the ages of the PMS stars using the $F814W_0$ vs. $(F814W-F160W)_0$ CMD in combination with the PARSEC 1.2S isochrones.

The full sample of 5404 PMS stars (cluster members detected in $F814W$ and $F160W$) has a mean age of $1.04 \pm 0.71$~Myr with $\sim60\%$ of all stars being between 1.0--2.0~Myr old. The full sample age is representative for the Wd2 cluster age (see Sect.~\ref{sec:stellar_ages}). The cluster age is also in good agreement with the age estimated by \citet[][1.5--2Myr]{Ascenso_07} and the theoretical MS lifetime of massive O stars of 2--5~Myr \citep[see Tab 1.1 in][]{Sparke_07}. Therefore, Wd2 has the same age or is even younger than other very young star clusters like \object{NGC~3603} \citep[1~Myr,][]{Pang_13}, \object{Trumpler~14} \citep[$\le 2$~Myr,][]{Carraro_04_Tr14} in the \object{Carina Nebula} \citep{Smith_08}, \object{R136} in the \object{Large Magellanic Cloud} \citep[1--4~Myr,][]{Hunter_95,Walborn_97,Sabbi_12}, \object{NGC~602} \citep{Cignoni_09} and \object{NGC~346} \citep{Cignoni_10ApJ} both in the SMC, or the \object{Arches} cluster \citep{Figer_02,Figer_05}. It is also younger than \object{Westerlund~1} ($5.0 \pm 1.0$~Myr), the most massive young star cluster known in the MW \citep{Clark_05,Gennaro_11,Lim_13}. Comparing the $F814W_0$ vs. $(F814W-F160W)_0$ CMDs of the four different regions MC, NC, the Wd2 cluster outskirts, and the periphery of RCW~49, we do not find any significant age difference between the regions (see Tab.~\ref{tab:spatial_distribution}). It appears that the MC and the NC are coeval.

Following the method applied in \citet{deMarchi_10} we used the individually extinction-corrected $F555W$, $F814W$, and $F658N$ photometry to select 240 H$\alpha$ excess emission stars in the RCW~49 region. We used the ATLAS9 model atmospheres \citep{Castelli_Kurucz_03} and the Stellar Spectral Flux Library by \citet{Pickles_98} to obtain interpolated $R$-band photometry from the $F555W_0$ and $F814W_0$ filters to get a reference template (see Appendix~\ref{sec:R_band}). Using TCDs we selected all stars as H$\alpha$ excess emission stars that are located at least $5\sigma$ above the continuum emission. Additionally, all stars must have an H$\alpha$ emission line $\rm{EW}>10\rm{\AA}$. A $(F555W-F814W)_0 > 0.2$~mag criterion is used to exclude possible Ae/Be candidates (see Sect. \ref{sec:Halpha_emission}). This yields 24 Ae/Be candidates (see Sect.~\ref{sec:AeBe_candiates}), mainly located in the TO and MS region of the $F814W_0$ vs. $(F814W-F160W)_0$ CMD  (see Fig.~\ref{fig:F814W-F160W_F814W}) and 240 H$\alpha$ excess emission stars with a mean H$\alpha$ luminosity $L(\rm{H}\alpha)=\left(1.67 \pm 0.45\right) \cdot 10^{-31}~\rm{erg}~\rm{s}^{-1}$ and a mass accretion rate of $\dot M_{\rm{acc}}=\left(4.43 \pm 1.68 \right) \cdot 10^{-8}~M_\odot~\rm{yr}^{-1}$. The mean age is $0.62 \pm 0.57$~Myr. The MC and NC host at least 36 and 26 H$\alpha$ excess emission stars, respectively, while the remaining part of Wd2 cluster contains at least 106. The remaining 72 are located in the periphery (see Tab.~\ref{tab:spatial_distribution}). The mean mass accretion rate in Wd2 is $\sim 70\%$ higher than in the SN~1987~A field \citep[$\dot M_{\rm{acc}}=2.6 \cdot  10^{-8}~M_\odot~\rm{yr}^{-1}$,][]{deMarchi_10}, $\sim 77\%$ higher than in NGC~602 \citep[$\sim 2.5 \cdot 10^{-8}~M_\odot~\rm{yr}^{-1}$,][]{deMarchi_13a}, and $\sim 14\%$ higher than in NGC~346 \citep[$3.9 \cdot 10^{-8}~M_\odot~\rm{yr}^{-1}$,][]{deMarchi_11a}. With a mean age of $\sim 1$~Myr Wd2 is younger than the PMS populations investigated by the other studies, which explains the higher mass accretion rate. Taking the younger age and the uncertainty range into account, the mass accretion rates determined in this paper are consistent with the theoretical studies of \citet{Hartmann_98} and the collected data of \citet{Calvet_00} for a number of star-forming regions. \citet{Hartmann_98} showed in their theoretical study of the evolution of viscous disks that the mass accretion rate decreases with increasing age ($\dot M \propto t^{-\eta}$). This was confirmed in many observational studies for different regions inside and outside the MW \citep[e.g.,][]{Calvet_00,Sicilia-Aguilar_06,Fang_09,deMarchi_13a}, yet the slope is poorly constrained. We analyzed our bona-fide sample of 240 mass-accreting stars and determined a decreasing slope of $\eta=0.44 \pm 0.04$, which is in agreement with other studies, taking into account the large uncertainty.

The FUV flux emitted by the luminous OB stars can lead to a shorter disk lifetime due to erosion \citep[e.g.,][]{Clarke_07}. \citet{Anderson_13} studied the effects of photoevaporation in the close vicinity (0.1--0.5~pc) of OB stars. Most of their disks were completely dispersed within 0.5--3.0~Myr. In our study of Wd2 we used the centers of the MC and NC and calculated the projected geometric center of all known OB stars within 0.5~pc (red crosses in Fig.~\ref{fig:Halpha_spatial_dist}). We then calculated the mean mass accretion rate in annuli of $15''$ or 0.3~pc going outwards from the respective centers (see Fig.~\ref{fig:radial_Halpha_dist}). The median mass accretion rate in the Wd2 cluster is $4.43~\cdot 10^{-8}~M_\odot \rm{yr}^{-1}$ and thus $\sim 25$--$30\%$ higher than in the MC ($3.32~\cdot 10^{-8}~M_\odot \rm{yr}^{-1}$) and NC ($3.12~\cdot 10^{-8}~M_\odot \rm{yr}^{-1}$). With increasing distance from the respective centers of the two density concentrations the mass accretion rate steeply increases by 60\% in the MC and 68\% in the NC within the innermost $30''$ (0.6~pc) and $45''$ (0.9~pc), respectively. With an increasing number of OB stars the mass accretion rate drops by 5--22\% (see Fig.~\ref{fig:radial_Halpha_dist}). Far away ($\gtrapprox 0.5$~pc) from the OB stars the mass accretion rate rises to a peak value of $5.9~\cdot 10^{-8}~M_\odot \rm{yr}^{-1}$. Despite the large uncertainty in the mass accretion rate, the effect of the increased rate of disk destruction is visible. This effect was also seen in other massive star-forming regions, e.g.,, by \citet{deMarchi_10} for the region around SN~1987~A and by \citet{Stolte_04} for NGC~3603 and supports the theoretical scenario of \citet{Clarke_07} and \citet{Anderson_13}.

In \citet{Zeidler_16c} we will provide completeness tests and a more sophisticated analysis of the spatial distribution of the stellar population in Wd2 than in Paper~I. Furthermore, we will determine the present-day mass function, as well as the mass of the Wd2 cluster as a whole and of its sub-clusters.

\acknowledgements
Based on observations made with the NASA/ESA Hubble Space Telescope, obtained at the Space Telescope Science Institute, which is operated by the Association of Universities for Research in Astronomy, Inc., under NASA contract NAS 5-26555. These observations are associated with program \#13038.

P.Z., E.K.G., and A.P. acknowledge support by Sonderforschungsbereich 881 (SFB 881, ''The Milky Way System'') of the German Research Foundation, particularly via subproject B5. M.T. has been partially funded by PRIN-MIUR 2010LY5N2T.

We thank ESA for the financial support for P.Z. to visit STScI for a productive scientific collaboration.

We thank the referee for the helpful comments to improve the quality of the paper.

\facility{HST (ACS, WFC3)}

\appendix
\section{The $R$-band interpolation}
\label{sec:R_band}
\begin{figure}[htb]
	\resizebox{\hsize}{!}{\includegraphics{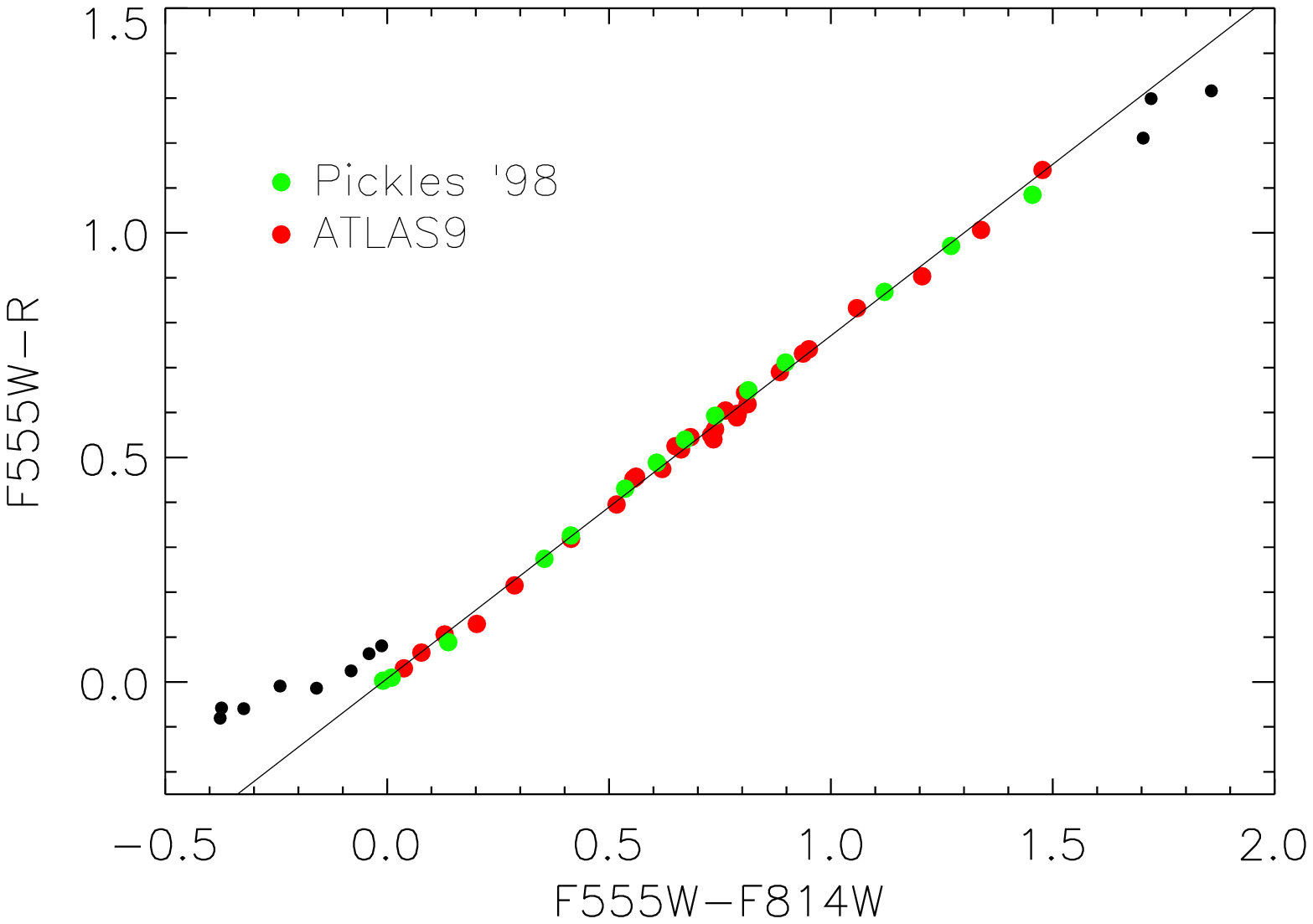}}
	\caption{The $F555W-R$ vs. $F555W-F814W$ TCD for the photometry derived from the ATLAS9 model atmospheres \citep[red points,][]{Castelli_Kurucz_03} and the Stellar Spectral Flux Library by \citet[][green points]{Pickles_98} for spectral types between A0V($T=9500$~K) and K5V($T=4250$~K). All spectral types between A2V--O5V and K7V--K5V are excluded because of their locations outside the linear regime of the detector response (black dots). The black line marks the least-squares linear fit.}
	\label{fig:R-band_interpolation}
\end{figure}

To better identify H$\alpha$ excess sources we combined the $F555W_0$ and $F814W_0$ photometry to produce an interpolated $R$-band. In order to study the relation of Johnson's $R$-band \citep{Johnson_53} and the ACS/WFC $F555W$ and $F814W$ filters \citep{ACS} we used the \textsc{synphot/calcphot} routine\footnote{Synphot is a product of the Space Telescope Science Institute, which is operated by AURA for NASA.}\citep{Synphot} in combination with the ATLAS9 model atmospheres \citep{Castelli_Kurucz_03} and the Stellar Spectral Flux Library by \citet{Pickles_98}. We determined the artificial stellar magnitudes by folding the respective filter curves with the stellar spectra for main-sequence stars (ATLAS9: K7V--A0V and \citet{Pickles_98}: M6V--O5V). In Fig.~\ref{fig:R-band_interpolation} we show the $F555W-R$ vs. $F555W-F814W$ TCD diagram. The red points are the photometry determined using the ATLAS9 models and the green data points are determined using the \citet{Pickles_98} library. The black data points, representing the spectral types of A2V--O5V and K7V--K5V, are excluded from the fit because the relation becomes non-linear.

For spectral types between A0V($T=9500$~K) and K5V($T=4250$~K) the photometric relation is remarkably linear. In this range we performed a least-squares linear fit. As a result we got

\begin{eqnarray}
\label{eq:R-band_interpolation}
R=0.237 \cdot F555W + 0.763 \cdot F814W -0.008,
\end{eqnarray}

with an uncertainty $\sigma=0.007$. This relation was then used to calculate the interpolated $R$-band photometry from the ACS $F555W_0$ and $F814W_0$ photometry.

\section{Calibration of the $F555W$ reddening correction}
\label{sec:F555W_calib}
The HST filters are just a rough representation of the Johnson-Cousins photometric system \citep{Johnson_53} and constitute their own photometric system \citep[see throughput curves of Fig.~13 of][]{Sirianni_05}. A detailed description and calibration cookbook for the HST/ACS filters are provided in \citet{Sirianni_05}. So far we always used the internal HST filter sets apart from the reddening correction via the color excess map $E(B-V)$. In \citet{Zeidler_15} we showed a detailed description of the transformation of $E(B-V)$ to any filter set $E(\lambda_1-\lambda_2)$ based on Cardelli's extinction law \citep{Cardelli_89}. This extinction law depends on the total-to-selective extinction parameter $R_V$ and assumes a different analytical form depending on the wavelength, divided into three wavelength regimes: infrared, optical/near-infrared, and ultraviolet. In the optical/near-infrared it is described as a seventh degree polynomial \citep[see equations 1, 3a, b of][]{Cardelli_89} that fits their five passbands ($UBVRI$).

We detected a discrepancy in the colors when we used the $F555W$ filter between the reddening-corrected photometric catalog and the theoretical PARSEC 1.2S isochrones \citep{Bressan_12}. We translated the color excess $E(B-V)$ to a total extinction $A(\lambda)$ at the pivot wavelength $\lambda$ for each of the used HST filters using the definition of the total-to-selective extinction $R_V=A_V/E(B-V)$ and equation~(1) of \citet{Cardelli_89}:

\begin{eqnarray}
\label{eq:A_lambda}
A\left(\lambda \right)=\left[a(x) \cdot R_V + b(x)\right] \cdot E(B-V).
\end{eqnarray}

\begin{figure*}[htb]
\resizebox{\hsize}{!}{\includegraphics{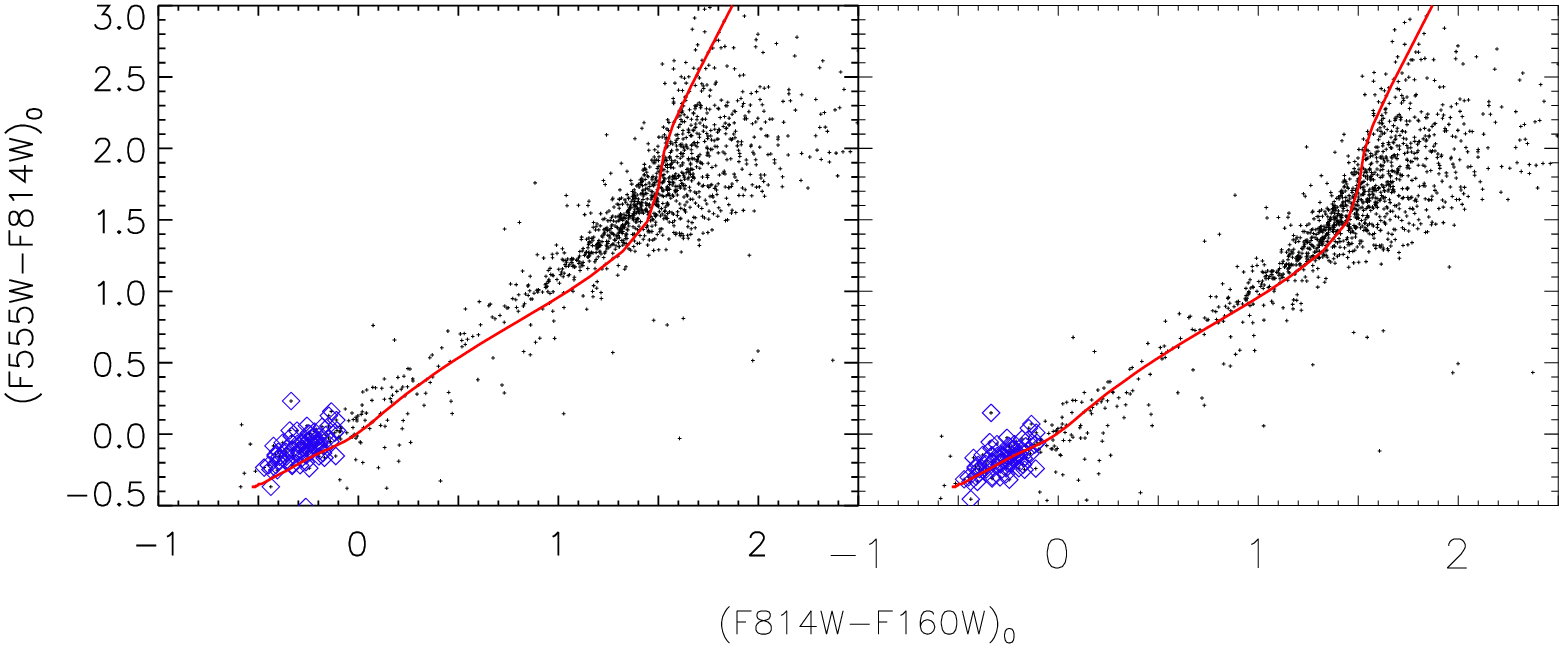}}
\caption{The reddening-corrected $(F555W-F814W)$ vs. $(F814W-F160W)$ TCD of the selected stars in RCW~49. \textbf{Left:} The dereddening was performed with a color-excess transformation $A(F555W)/A(V)=1.024$ using the pivot wavelength of the $F555W$ filter. \textbf{Right:} We used a color-excess transformation $A(F555W)/A(V)=1.038$ for the dereddening to fit the MS to the ZAMS in the TCDs. The blue diamonds are the selected MS stars used for the fit. The red line represents the ZAMS from the PARSEC 2.1S models.}
\label{fig:F555W_comparison}
\end{figure*}

$x=1/\lambda~[\mu\rm{m}^{-1}]$ while $a(x)$ and $b(x)$ are the inverse wavelength-dependent coefficients of Cardelli's extinction law \citep{Cardelli_89} at the pivot wavelength of the HST filters \citep[see Table~6,][]{Zeidler_15}. The $F555W$ filter is the only filter in our observations whose pivot wavelength of $\lambda_P=536.1$~nm is bluer than Johnsons-Cousin's $V$-band, while the width is larger than the V-band width \citep[see Fig.~1,][]{Maiz_Apellaniz_13}. The pivot wavelength is a weighted mean taking into account the filter's throughput curve. The extinction law is just evaluated at one point. This fact is also mentioned by \citet{Maiz_Apellaniz_13} and \citet{Sirianni_05}. At the location of the $V$-band the inverse wavelength-dependent coefficient $b(x)$ of Cardelli's extinction law \citep{Cardelli_89} changes its sign and so the evaluation of the $A(F555W)/A(V)$ at just $\lambda_P=536.1$~nm can cause errors. In our case this leads to an under-correction of the reddening for the $F555W$ filter. In the left panel of Fig.~\ref{fig:F555W_comparison} we give the example of the reddening-corrected $(F555W-F814W)$ vs. $(F814W-F160W)$ TCD.

To correct $A(F555W)/A(V)$ we used four TCDs based on the $F814W$, $F658N$, $F125W$, and $F160W$ filters. We selected the MS stars and fitted them simultaneously to the ZAMS by adjusting $A(F555W)/A(V)$ taking into account the photometric errors. It is possible to reduce this problem to a linear fit of the following form:

\begin{eqnarray}
\label{eq:F555W_correction}
\left(F555W-\mathrm{X}_0\right)-\left(F555W-\mathrm{X}\right)_{\mathrm{ZAMS}} \nonumber \\
=\frac{A(F555W)}{A(V)} \cdot E(B-V).
\end{eqnarray}

X represents the different filters. In Fig.~\ref{eq:F555W_correction} the relations for four different filters are plotted including the overall best fit, which results in $A(F555W)/A(V)=1.038$. This implies an increase of 1.4\% for the ratio $A(F555W)/A(V)$ with a total-to-selective extinction of $R_V=3.95$. As an example and comparison, we give in the right panel of Fig.~\ref{fig:F555W_comparison} the reddening-corrected $(F555W-F814W)$ vs. $(F814W-F160W)$ TCD for the adjusted $A(F555W)/A(V)$ value.

\begin{figure}[htb]
\resizebox{\hsize}{!}{\includegraphics{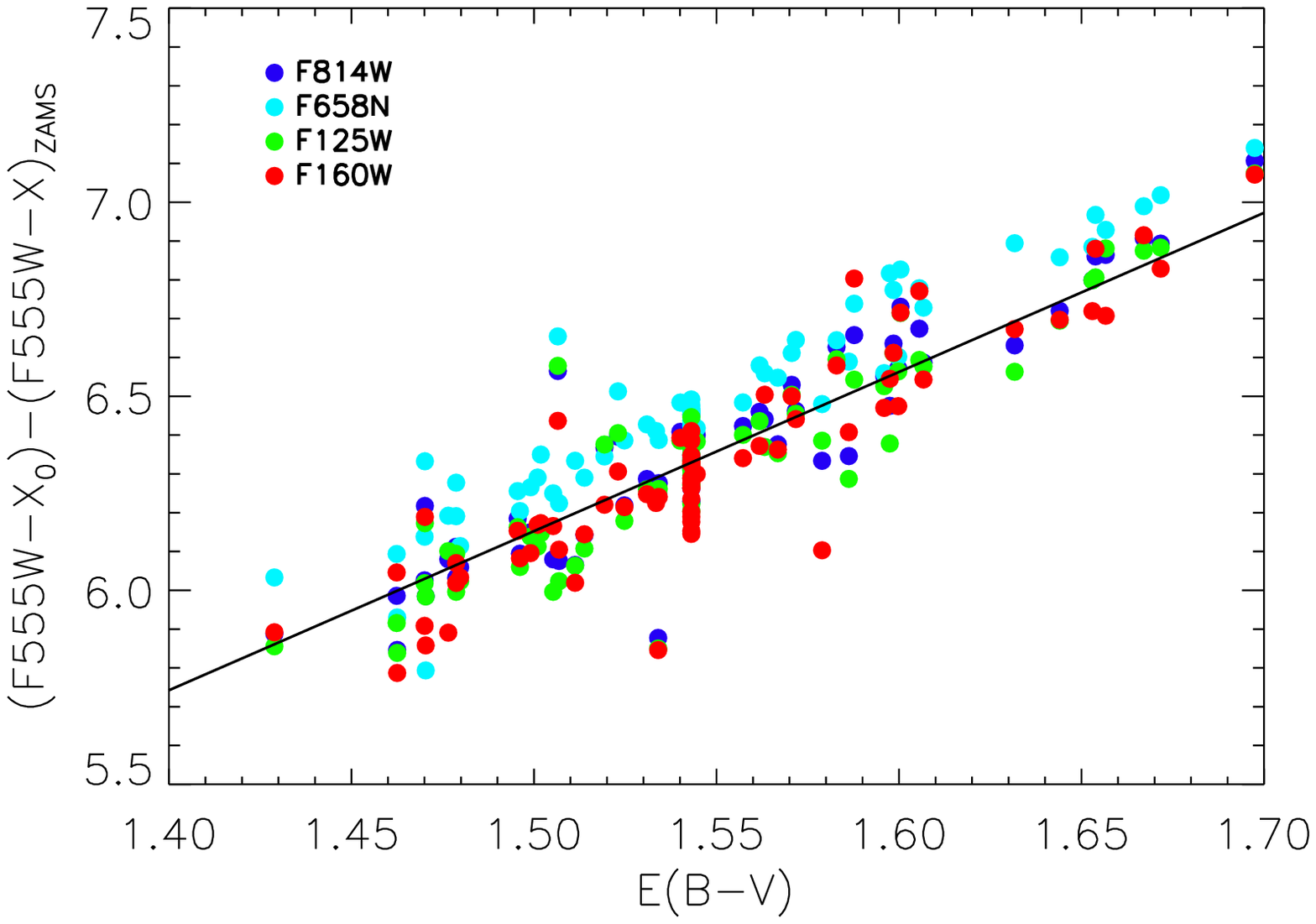}}
\caption{The linear fit for the reddening correction using all main-sequence stars for the four TCDs involving the $F555W$ band. The straight line represents the best fit taking into account photometric errors.}
\label{fig:F814W_correction}
\end{figure}

\newpage
\section{Tables}

\floattable
\onecolumngrid
\begin{deluxetable}{lcrrrrrrrrrrr}
\tabletypesize{\scriptsize}
	\tablecaption{The age distribution of the PMS and H$\alpha$ excess sources\label{tab:ages}}
	\tablehead{
		\multicolumn{1}{c}{Source} &\multicolumn{1}{c}{Panel} &\multicolumn{2}{c}{0.1~Myr} & \multicolumn{2}{c}{0.25~Myr} &  \multicolumn{2}{c}{0.5~Myr} & \multicolumn{2}{c}{1.0~Myr} & \multicolumn{2}{c}{2.0~Myr} & \multicolumn{1}{c}{Total}
	}
	\startdata
	Full-sample PMS stars				  &	  & 547 & (10.1\%) & 550 &  (10.2\%) & 1081 &  (20.0\%) & 1552 &  (28.7\%) & 1674 &  (31.0\%) & 5404 \\
	\hspace{+1em} Main cluster		 	  &	a & 18  &  (3.6\%) & 44  &  (10.7\%) & 107  &  (21.5\%) & 144  &  (28.9\%) & 185  &  (35.3\%) & 498 \\
	\hspace{+1em} Northern clump 		  &	b & 9   &  (2.9\%) & 22  &  (7.1\%)  & 70   &  (22.6\%) & 101  &  (32.6\%) & 108  &  (34.8\%) & 310 \\
	\hspace{+1em} Westerlund 2			  &	c & 99  &  (5.4\%) & 155 &  (7.3\%)  & 366  &  (19.8\%) & 635  &  (34.4\%) & 589  &  (33.1\%) & 1844 \\
	\hspace{+1em} Periphery 			  &	d & 421 & (15.4\%) & 329 &  (12.0\%) & 538  &  (19.6\%) & 672  &  (24.5\%) & 792  &  (28.5\%) & 2752 \\[0.1cm]	
	Reduced-sample PMS stars 			  &	  & 192 &  (11.0\%) & 242 &  (16.8\%) & 485 &  (30.5\%) & 440 &  (25.5\%) & 331 &  (16.2\%) & 1690 \\
	\hspace{+1em} Main cluster			  &	a & 15  &   (5.7\%) & 35  &  (13.3\%) & 80  &  (30.4\%) & 79  &  (30.0\%) & 54  &  (20.6\%) & 263 \\
	\hspace{+1em} Northern clump 		  &	b & 5   &   (3.5\%) & 14  &  (9.9\%)  & 53  &  (37.3\%) & 37  &  (26.1\%) & 33  &  (23.2\%) & 142 \\
	\hspace{+1em} Westerlund 2			  &	c & 60  &   (8.9\%) & 108 &  (16.0\%) & 203 &  (30.2\%) & 190 &  (28.2\%) & 112 &  (16.7\%) & 673 \\
	\hspace{+1em} Periphery 			  &	d & 112 &  (19.9\%) & 85  &  (13.9\%) & 149 &  (24.3\%) & 134 &  (21.9\%) & 132 &  (20.0\%) & 612 \\[0.1cm]	
	H$\alpha$-excess &	  & 54 &  (22.5\%) &  49 &  (20.4\%) &  66 &  (27.5\%) &  45 &  (18.8\%) &  26 &  (10.8\%) & 240\\	
	\enddata
	\tablecomments{For each age bin we give the number of sources and in brackets the fraction of sources compared to the total number of objects. For each sample we also list the distribution within the sub-region described in (Paper~I and \cite{Zeidler_16c}). Column 2 (panel) gives the letter denoting the panel of the region in Fig.~\ref{fig:area_CMDs}.}
\end{deluxetable}
\twocolumngrid

\floattable
\onecolumngrid
\begin{deluxetable}{lrrrrrrrrr}
\tabletypesize{\scriptsize}
	\tablecaption{The different areas of RCW~49 \label{tab:spatial_distribution}}
	\tablehead{
		\multicolumn{1}{c}{\textsc{ }} &\multicolumn{2}{c}{\textsc{MC (a)}} &\multicolumn{2}{c}{NC (b)} & \multicolumn{2}{c}{Wd 2 (c)} &  \multicolumn{2}{c}{Periphery (d)} & \multicolumn{1}{c}{Total}
	}
	\startdata
	Full-sample PMS stars			 & 498 & (9.2\%) & 310 &  (5.7\%)  & 1844 &  (34.2\%) & 2752 &  (50.9\%) & 5404 \\
	\hspace{+1em} Mean age	 [Myr]	 & \multicolumn{2}{c}{$1.17 \pm 0.69$} & \multicolumn{2}{c}{$1.16 \pm 0.67$} & \multicolumn{2}{c}{$1.11 \pm 0.67$} & \multicolumn{2}{c}{$0.96 \pm 0.73$} &   \\	
	Reduced-sample PMS stars 			  & 263 & (15.6\%) & 142 &  (8.4\%)  & 673 &  (39.8\%) & 612 &  (36.2\%) & 1690 \\
	\hspace{+1em} Mean age	 [Myr]		     & \multicolumn{2}{c}{$0.90 \pm 0.63$} & \multicolumn{2}{c}{$0.94 \pm 0.64$} & \multicolumn{2}{c}{$0.82 \pm 0.61$} & \multicolumn{2}{c}{$0.83 \pm 0.69$} &   \\
	H$\alpha$-excess stars & 36 & (15.0\%) & 26 &  (10.8\%)  & 106 &  (44.2\%) & 72 &  (30.0\%) & 240 \\
	\hspace{+1em} Mean age	 [Myr]		     & \multicolumn{2}{c}{$0.64 \pm 0.59$} & \multicolumn{2}{c}{$0.77 \pm 0.61$} & \multicolumn{2}{c}{$0.63 \pm 0.55$} & \multicolumn{2}{c}{$0.54 \pm 0.58$} &   \\
	\hspace{+1em} Mean $\dot M$	$[10^{-8} M_\odot \rm{yr}^{-1}]$ 	         & \multicolumn{2}{c}{$3.32$} & \multicolumn{2}{c}{$3.12$} & \multicolumn{2}{c}{$4.84$} & \multicolumn{2}{c}{$5.70$} & $4.43$ \\
	\enddata
	\tablecomments{In this table we present a summary of the different properties of the stellar population in the different regions of RCW~49. The letters in brackets are the panel numbers in Fig.~\ref{fig:area_CMDs}.}
\end{deluxetable}
\twocolumngrid

\bibliographystyle{aasjournal}
\bibliography{Wd2_submission}

\end{document}